\documentclass[aps,prl,floats,floatfix,twocolumn,fleqn,linenumbers,longbibliography]{revtex4-1}

\RequirePackage{lineno} 
\renewcommand\thelinenumber{}

\usepackage{graphicx}
\usepackage{epstopdf}
\graphicspath{{./Figs/}} 

\usepackage{latexsym}
\usepackage{amsmath}
\usepackage{amssymb}
\usepackage{bm}
\usepackage{wasysym}

\usepackage{ifthen}
\usepackage{color}
\usepackage[colorlinks,allcolors=blue,bookmarks=true]{hyperref}

\newcommand{\trc}{\mbox{trace}}

\newcommand{\im}{\mbox{Im}}
\newcommand{\re}{\mbox{Re}}
\newcommand{\mass}{\mathsf{m}}

\newcommand{\lvec}[1]{{\reflectbox{\ensuremath{\vec{\reflectbox{\ensuremath{#1}}}}}}}

\newcommand{\bra}[1]{\left\langle #1 \right|}
\newcommand{\ket}[1]{\left| #1 \right\rangle}
\newcommand{\braket}[1]{\left\langle #1 \right\rangle }
\newcommand{\avg}[1]{\left\langle #1 \right\rangle }
\newcommand{\kb}[2]{ | #1 \rangle \langle #2 |}

\newcommand{\BraKet}[3]{\left\langle #1 \middle| #2 \middle| #3 \right\rangle}

\newcommand{\id}{1\!\!1}

\newcommand{\var}{\mbox{\text{Var}}}

\newcommand{\beq}{\begin{eqnarray}}
\newcommand{\eeq}{\end{eqnarray}}

\newcommand{\hide}[1]{}  
\newcommand{\rmrk}[1]{#1}    

\newcommand{\Eq}[1]{\textcolor{blue}{{equation}\!~(\ref{#1})}} 
 
\newcommand{\Fig}[1] {{\textcolor{blue}{Fig.}}~\!\!\ref{#1}}
\newcommand{\sect}[1]{{\bf #1.-- }}

\newcommand{\Cn}[1]{\begin{center} #1 \end{center}}
\newcommand{\hrefl}[1]{\href{#1}{[link]}}
\newcommand{\+}{^{\dag}}
\newcommand{\bdag}{{\bm{\dag}}}

\renewcommand{\thesection}{\arabic{section}}
\renewcommand{\thesubsection}{\arabic{subsection}}

\newcommand{\sectA}[1]
{
\addtocounter{section}{1}
\setcounter{subsection}{0}
\ \\
\pdfbookmark[1]{\thesection. \ #1}{sect.\thesection}
{\Large\bf $=\!=\!=\!=\!=\!=\;$ [\thesection] \ #1}
\nopagebreak
\vspace*{3mm}
}
\renewcommand{\section}{\sectA}

\newcommand{\sectB}[1]
{
\addtocounter{subsection}{1}
\ \\
\pdfbookmark[2]{\ \ \ \ \thesection.\thesubsection. \ #1}{subsect.\thesection.\thesubsection}
{\bf $=\!=\!=\!=\!=\!=\;$ [\thesection.\thesubsection] \ #1}
\nopagebreak
}
\renewcommand{\subsection}{\sectB}

\begin{document}

\title{Quantum stochastic transport along chains}

\author{Dekel Shapira, Doron Cohen}

\affiliation{
\mbox{Department of Physics, Ben-Gurion University of the Negev, Beer-Sheva 84105, Israel} 
}

\hide{
\begin{abstract}
We study the {\em spreading} along an infinite tight-binding chain, and the {\em relaxation} within a finite ring (chain with periodic boundary conditions). Specifically we address the interplay of {\em coherent} and {\em stochastic} transitions within the framework of an Ohmic master equation,  which leads to a non-monotonic dependence of the current on the bias. With added disorder it becomes the quantum version of the Sinai-Derrida-Hatano-Nelson model, which features sliding and delocalization transitions. We highlight counter-intuitive enhancement of disorder due to coherent hopping.     
\end{abstract}
}

\maketitle

\noindent {\Large\bf\textsl{Abstract}} 
\vspace*{1mm}

The {\em spreading} of a particle along a chain, and its {\em relaxation}, are central themes in statistical and quantum mechanics. One wonders what are the consequences of the interplay between {\em coherent} and {\em stochastic} transitions. This fundamental puzzle has not been addressed in the literature, though closely related themes were in the focus of the Physics literature throughout the last century, highlighting quantum versions of Brownian motion. Most recently this question has surfaced again in the context of photo-synthesis.  
Here we consider both an infinite tight-binding chain and a finite ring within the framework of an Ohmic master equation. With added disorder it becomes the quantum version of the Sinai-Derrida-Hatano-Nelson model, which features sliding and delocalization transitions. We highlight non-monotonic dependence of the current on the bias, and a counter-intuitive enhancement of the effective disorder due to coherent hopping.

\vspace*{5mm}
\noindent {\Large\bf\textsl{Introduction}} 
\vspace*{1mm}

A prototype problem in Physics is the dynamics of a particle along chain that consists of sites. 
If the dynamics is coherent one expects to observe ballistic motion and Bloch oscillations \cite{Hartmann_korsch_2004}, while for stochastic dynamics one expects to see diffusion and drift.
In the presence of disorder, additional fascinating effects emerge: an Anderson localization transition in the coherent problem, and a Sinai-Derrida sliding transition in the stochastic problem.
In practical applications the particle can be an exiton 
\cite{dubin2006macroscopic,nelson2018coherent,dekorsy2000coupled}.
Past literature regarding quantum spreading in chains,  
\cite{MadhukarPost1977,Weiss1985,Kumar1985,Dibyendu2008,Amir2009,lloyd2011quantum,Moix_2013,CaoSilbeyWu2013,Kaplan2017ExitSite,Kaplan2017B}, 
including publications that address the photo-synthesis theme 
\cite{amerongen2000photosynthetic,ritz2002quantum,FlemingCheng2009,plenio2008dephasing,Rebentrost_2009,Alan2009,Sarovar_2013,higgins2014superabsorption,celardo2012superradiance,park2016enhanced}, 
were focused mainly on the question how noise and dissipation affect coherent transport. 
\rmrk{In a sense, our interest is in the reversed question.}

\rmrk{In the present work we assume {\em independent} mechanisms for stochastic asymmetric (dissipative) transitions, and for coherent hamiltonian (conservative) transitions. Such setup is not common: the standard models do not allow to tune on and off the two mechanisms independently. The question arises how the two mechanisms affect each other. Can we simply ``sum up" known results for stochastic transport with known results for coherent motion in noisy environment? We shall see that the answer is not trivial. The main surprises come out once we take into account the presence of disorder (see below). An optional way to phrase the question: what is the quantum version of the prototype stochastic problem that is known in the literature as {\em random walk in random environment}. As we know from the above cited works, due to disorder, the stochastic dissipative dynamics is not merely a simple minded Brownian motion. We would like to know whether coherence has any implication on the predicted disorder-related crossovers.}

\begin{figure}[b]
\centering
\includegraphics[]{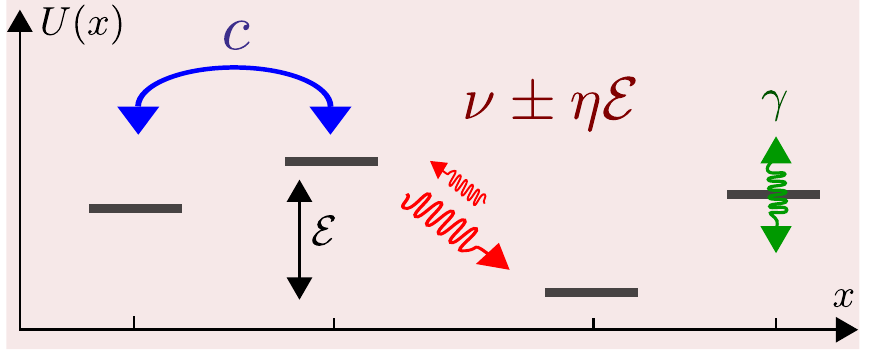} \\
\caption{\label{fig:schematic-transitions}
{\bf (1) Illustration of the model system.}
Each site of the chain is represented by a line segment positioned according to its $x$ coordinate and potential $U(x)$.
Blue arrow labeled by $c$ represents the possibility for a coherent hopping  between two sites.
Red arrows represent bath induced stochastic transitions between two sites.
The local bath that is responsible for the latter fluctuates with intensity $\nu$, and  the induced transitions are asymmetric if $\eta=\nu/(2T)$ is non-zero (finite temperatures).
Note that their ratio is $\exp \left({-\mathcal{E}/T} \right)$ in leading order.
The green wiggle lines represent a local bath that induces fluctuations of intensity $\gamma$ of the on-site potential. 
Without the baths it is the Anderson model for coherent transport and localization in disordered chain. In the other extreme, if only the stochastic transitions are present, it is the Sinai-Derrida model  for motion in random environment. The latter exhibits a sliding transition as the bias is increased,
and an associated Hatano-Nelson delocalization transition once relaxation in a closed ring is considered.}
\end{figure}

We consider a chain whose sites are labeled by~$x$. The particle, or the exiton, can move from site to site (near neighbor transitions only). The transitions are determined by two major parameters: the hopping frequency~($c$) that controls the coherent hopping; and the fluctuations intensity~($\nu$) that controls the environmentally-induced stochastic transitions. 
At finite temperature~$T$ there is also a dissipation coefficient ${\eta=\nu/(2T)}$ that is responsible for the asymmetry of the stochastic transitions. 
On top we might have bias~($\mathcal{E}$), on-site noisy fluctuations~($\gamma$), and different types of disorder. 
The model is illustrated in \Fig{fig:schematic-transitions}. 

The dynamics is governed by a master equation for the probability matrix
\beq \label{e1}
\frac{d\rho}{dt} \ = \ \mathcal{L} \rho \ = \ 
-i[\bm{H}^{(c)},\rho] + \left( \mathcal{L}^{(\text{B})} + \mathcal{L}^{(\text{S})} \right) \rho
\eeq
where the dissipators $\mathcal{L}^{(\text{B})} \propto \nu$ and  $\mathcal{L}^{(\text{S})} \propto \gamma$ are due to the interaction with the environment. They are responsible for the stochastic aspect of the dynamics. 
The Hamiltonian $\bm{H}^{(c)}$ contains an on-site potential $U(x)$, and a sum over hopping terms  $(c/2) \kb{x \pm 1}{x}$. Accordingly it takes the form
\beq \label{e02}
\bm{H}^{(c)} \ \ \equiv \ \ U(\bm{x}) -c \cos(\bm{p})
\eeq
where $\bm{p}$ is the momentum operator. 
The unit of length is the site spacing ($x$~is an integer), 
and the field is 
\beq
\mathcal{E}_x \ \equiv \ -\left( U(x{+}1)-U(x) \right)
\eeq

In the absence of stochastic terms, coherent transport in ordered chain leads to ballistic motion (without bias) and exhibits Bloch-oscillations (with bias). In disordered chain the spreading is suppressed due to Anderson-localization. The effect of noise and dissipation on coherent transport due to $\mathcal{L}^{(\text{S})}$ has been extensively studied. In the Caldeira-Leggett model \cite{Caldeira1983,CALDEIRA1983374} the interaction is with homogeneous fluctuating environment, leading to Brownian motion with Gaussian spreading. If the interaction is with non-homogeneous fluctuating environment (short spatial correlation scale) the spreading is the sum of a decaying {\em coherent Gaussian} and a scattered {\em Stochastic Gaussian} \cite{Cohen1997}. The tight binding version of this model has been studied in \cite{EspositoGaspard2005}. It has been found that the decoherence and the stochastic-like evolution are dictated by different bands of the Lindblad $\mathcal{L}$-spectrum that correspond, respectively, to the dephasing and to the relaxation rates in NMR studies of two-level dynamics.

\begin{figure}
\centering
\includegraphics[]{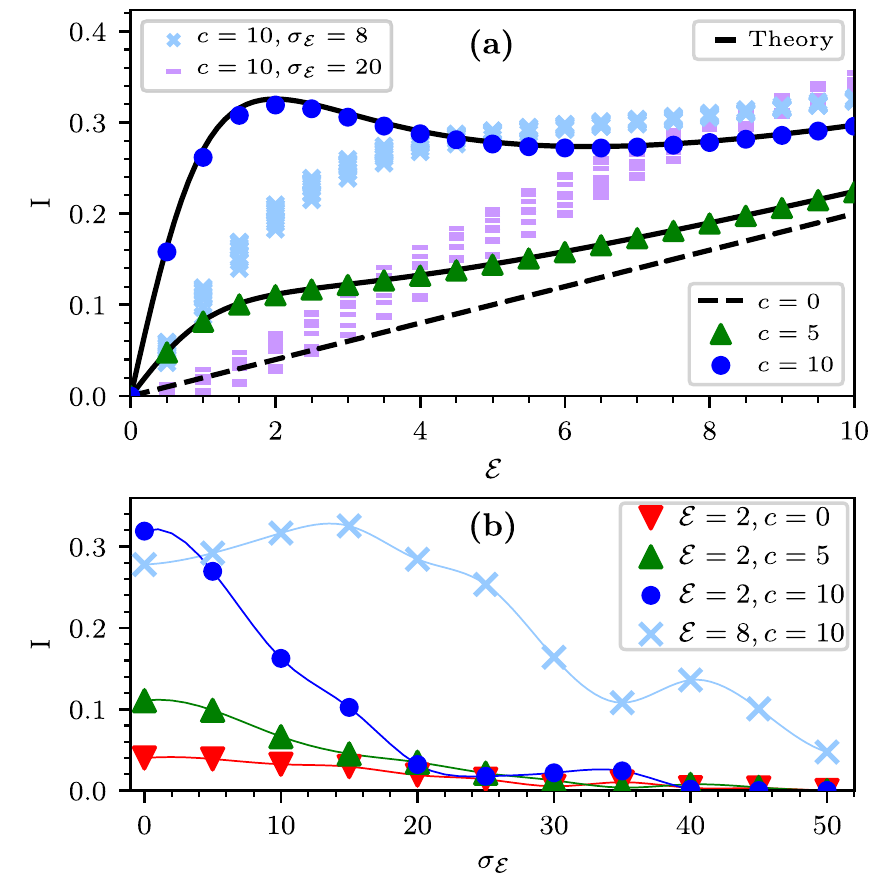}
\caption{\label{f1}
{\bf The NESS current for a biased chain, with and without disorder.}
\textit{(a)} The NESS current as a function of $\mathcal{E}$. 
Black lines are based on \Eq{eq:I} for clean system with ${c=0,5,10}$ and $\nu=1$ and $\eta=0.01$. 
Symbols are based on numerical determination of the NESS for a ring of $L{=}500$ sites. 
We display 10 independent realizations of the disorder for each value of disorder strength $\sigma_{\mathcal{E}}$, 
while ${c{=}10}$ is kept the same. 
\textit{(b)} The average NESS current as a function of $\sigma_{\mathcal{E}}$ 
for $\mathcal{E}{=}2$. In the $c{=}10$ case also for $\mathcal{E}{=}8$.
Thin-lines are a guide to the eye.
}
\end{figure}

In the other extreme of purely stochastic dynamics, ignoring quantum effects, the disordered model, aka {\em random walk in random environment}, has been extensively studied by Sinai, Derrida, and followers  \cite{Dyson1953,Sinai1983,DerridaPomeau1982,Derrida1983,havlin1987diffusion,Bouchaud1990,BOUCHAUD1990a}. Without bias the spreading becomes sub-diffusive, while above some critical bias the drift-velocity becomes finite, aka {\em sliding transition}. Strongly related is the transition from over-damped to under-damped relaxation that has been studied for a finite-size ring geometry \cite{HurowitzCohen2016,HurowitzCohen2016a}. The latter involves {\em delocalization transition} that has been highlighted for non-hermitian Hamiltonians in the works of Hatano, Nelson and followers \cite{Hatano1996,Hatano1997,Hatano1998,LubenskyNelson2000,LubenskyNelson2002,Amir2015,Amir2016,Shnerb99}.

One should realize that the two extreme limits of coherent and stochastic spreading have to be bridged 
within the framework of a model that includes an $\mathcal{L}^{(\text{B})}$ term, 
not just an $\mathcal{L}^{(\text{S})}$ term. 
Furthermore, a proper modeling requires the distinction between two types of Master equations. In one extreme we have the {\em Pauli version}. Traditionally this version is justified by the secular approximation that assumes weak system-bath interaction. In the other extreme we have the {\em Ohmic version} that assumes short correlation time. The so called ``singular coupling limit" can be regarded as an optional way to formalize the short correlation time assumption \cite{Rivas2012}. 
Clearly in the mesoscopic context it is more appropriate to adopt the Ohmic version, and regard the Pauli version of the dissipator as a formal approximation. 

\begin{figure}
\centering
\includegraphics[]{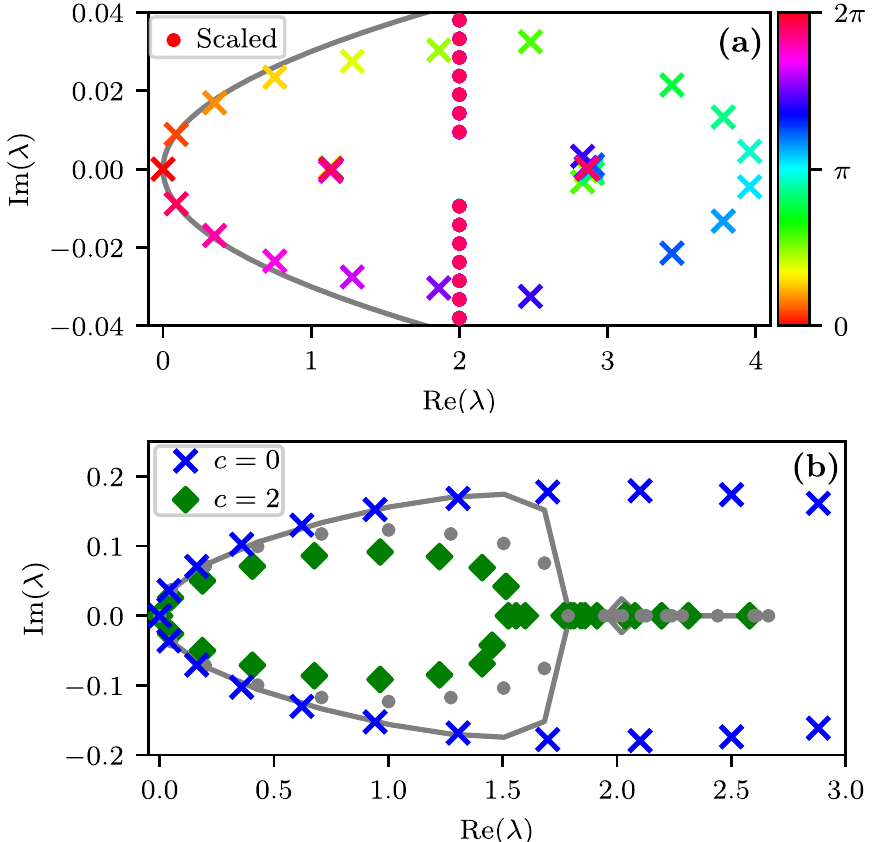}
\caption{\label{f2} 
{\bf The Lindblad $\mathcal{L}$-spectrum for both non-disordered and disordered rings.}
\textit{(a)}~ 
The spectrum for a non-disordered ring of ${L=21}$ sites. 
The eigenvalues that form an ellipse correspond to the stochastic-like relaxation modes. 
The eigenvalues that bunch together at $\lambda \sim {1,3}$
are the $\mp$ over-damped decoherence modes. The other eigenvalues
along $\re(\lambda)=2$ correspond to under-damped decoherence modes 
(each point is in a fact a band of $L$ overlapping eigenvalues).
The dashed gray-line is based on \Eq{eVD}.
For presentation purpose the eigenvalues marked with dot are scaled by a factor of $0.001$ along the vertical axis. 
The colors indicate the $q$ of each eigenvalue.
The parameters are $\nu{=}1, \eta{=}0.01, c{=}0.1, \mathcal{E}{=}0.5$.
\textit{(b)}~ 
The relaxation spectrum with disorder (decoherence modes are excluded).
The spectrum for a chain with a given disorder is displayed, once with $c{=}0$ and once with $c{=}2$.
The gray-circles and the gray-line are the three-band and one-band approximations for $c{=}2$.
The other parameters are $L{=}31,\mathcal{E}{=}3,\sigma_{\mathcal{E}}{=}1.5, \nu{=}1, \eta{=}0.03$.
}
\end{figure}

\sect{Outline}
The model is presented in terms of an Ohmic master equation. 
The units of time are chosen such that the basic model parameters are 
$(c,\mathcal{E},\nu{\equiv}1,\eta)$ and the strength of the disorder $\sigma_{\mathcal{E}}$.
The interest is in the diffusion coefficient~$D$, 
the $\mathcal{E}$-induced drift velocity~$v$, 
the implied non equilibrium steady state (NESS) current~$I \equiv (1/L)v$ for a ring of length $L$, 
and the associated Lindblad \mbox{$\mathcal{L}$-spectrum}. The latter is determined via $\mathcal{L} \rho = - \lambda \rho$, 
which provides both the relaxation-modes and the decoherence-modes.   
In particular we observe that the NESS current depends non-monotonically on the bias (\Fig{f1}a), 
and that surprisingly it can be enhanced by disorder (\Fig{f1}b).
In a disordered ring, counter-intuitively, 
relaxation modes become over-damped if coherent transitions are switched on (\Fig{f2}).

\vspace*{5mm}
\noindent {\Large\bf\textsl{Results}} 
\vspace*{1mm}

\sect{The model}
The isolated chain is defined by the~$\bm{H}^{(c)}$ Hamiltonian \Eq{e02}, 
that describes a particle or an exiton that can hop along a one-dimensional chain 
whose sites are labeled by~$x$.
The field ${\mathcal{E}_x}$ might be non-uniform.  
For the average value of the field we maintain the notation~$\mathcal{E}$, 
while the random component is distributed uniformly (box distribution)
within ${ [-\sigma_{\mathcal{E}}, \sigma_{\mathcal{E}}] }$.  
We regard each pair of neighboring sites as a two-level system \cite{SM1}. Accordingly we distinguish between two types of terms in the master equation: those that originate from temporal fluctuations of the potential (dephasing due to noisy detuning), and those that are responsible to stochastic transition between the sites (incoherent hopping). The latter are implied by the replacement ${(c/2)\mapsto (c/2)+ f(t)}$ at the pertinent bonds, where $f(t)$ is a bath operator that is characterized by fluctuation intensity~$\nu$, and temperature~$T$. Hence the system bath coupling term is $-\bm{W}_xf(t)$, where ${\bm{W}_x = (\bm{D}_x + \bm{D}_x^{\dag})}$ and ${\bm{D}_x = \kb{x{+}1}{x}}$. The baths of different bonds are uncorrelated, accordingly the bond-related dissipator takes the form 
\beq \label{e2}
\mathcal{L}^{(\text{B})} \rho = 
-\sum_{x} \left(
\dfrac{\nu}{2} [\bm{W}_x, [\bm{W}_x, \rho]] 
+ \dfrac{\eta}{2}\, i [\bm{W}_x, \{\bm{V}_x, \rho\}]  
\right) \ \ 
\eeq
where ${\eta=\nu/(2T)}$ is the friction coefficient, and  
\beq \label{eFR}
\bm{V}_x \ \equiv \ i[\bm{H}^{(c)}, \bm{W}_x]
\eeq 
The friction terms represent the response of the bath 
to the rate of change of the $\bm{W}_x$. 
Note that for getting the conventional Fokker-Planck equation
the system-bath coupling term would be $-{\bm{x}} f(t)$, 
and $\bm{V}$ would become the velocity operator. Here we assume 
interaction with local baths that in general might have 
different temperatures.
See Methods for some extra technical details
regarding the master equation, the nature of the disorder,  
and the handling of the periodic boundary 
conditions for the ring configuration.

\sect{Pauli-type dynamics}
For pedagogical purpose let us consider first a uniform non-disordered ring without coherent hopping. 
Furthermore, let us adopt the simplified Pauli-like version of the dissipator (see Methods). 
Consequently the dynamics of the on-site probabilities ${p_x \equiv \rho_{x,x}}$  
decouples from that of the off-diagonal terms.
Namely, one obtains for the probabilities a simple rate equation, 
where the transition rates between sites are  
\beq \label{e5}
w^{\pm} \ = \ \nu \pm \eta \mathcal{E} 
\eeq
in agreement with Fermi-golden-rule (FGR). 
Note that in leading order $[w^{-}/w^{+}] \approx \exp(-\mathcal{E}/T)$ 
as expected from detailed balance considerations. 
It follows that the drift velocity and the diffusion coefficient are: 
\beq \label{eVP}
v \ &=& \ (w^{+} - w^{-}) \ = \ 2\eta \mathcal{E} 
\\  \label{eDP}
D \ &=& \ \frac{1}{2}(w^{+} + w^{-}) \ = \ \nu  
\eeq
Consequently one finds two distinct sets of modes: the stochastic-like relaxation modes 
that are implied by the rate equation for the probabilities, 
and off-diagonal decoherence modes.  
The latter share the {\em same} decay rate ${\gamma_0 = w^{+} + w^{-} + \gamma}$, 
where $\gamma$ stands for optional extra off-diagonal decoherence due to on-site fluctuations.
An evolving wavepacket \cite{SM3} will decompose into coherent decaying component that is suppressed by factor $e^{-\gamma_0 t}$, 
and an emerging stochastic component that drifts with velocity $v$ and diffuse with coefficient $D$.

\sect{Full Ohmic treatment}
The state of the particle in the standard representation is given
by ${\rho_x(r) \equiv \BraKet{x}{\rho}{x+r}}$. 
The master equation, \Eq{e1} with \Eq{e2}, couples  
the dynamics of the on-site probabilities ${p_x \equiv \rho_x(0)}$    
to that of the off-diagonal elements ${\rho_x(r \ne 0)}$.   
The generator $\mathcal{L}$ can be written as a sum of several terms \cite{SM5}: 
\beq \label{e4}
\mathcal{L} = 
\mathcal{E} \mathcal{L}^{(\mathcal{E})}
+c\mathcal{L}^{(c)}
+\nu \mathcal{L}^{(\nu)}
+\eta \mathcal{E} \mathcal{L}^{(\tilde{\mathcal{E}})}
+\eta c \mathcal{L}^{(\tilde{c})}
\eeq
Each term is a super-matrix that operates on the super vector ${\rho_x(r)}$.
The first two terms $\mathcal{L}^{(c)}$ and $\mathcal{L}^{(\mathcal{E})}$ arise from the Hamiltonian \Eq{e02}.
The $\mathcal{L}^{(\nu)}$ term arise from the first term of \Eq{e2}, 
which represent noise-induced transitions. 
The remaining two friction-terms (proportional to $\eta$) 
arise from \Eq{eFR}, and correspond to the two terms in the Hamiltonian.

\begin{figure}
\includegraphics{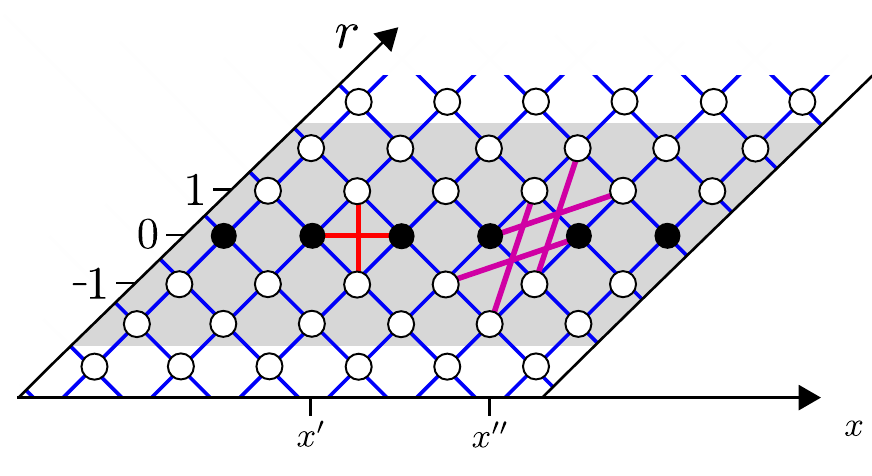}                                                      %
\caption{\label{fig:schematic-transitions-all}
{\bf Diagrammatic representation of the couplings in the master equation.}
A diagonal strip of the probability matrix  $\rho_x(r)$  is illustrated. 
The diagonal elements $p_x=\rho_x(0)$ are represented by filled circles, and the off-diagonal  terms by empty circles.
The Lindblad generator $\mathcal{L}$ induces ``transitions" between the elements. 
The blue grid lines indicate $c$-induced couplings. 
The other couplings within $|r| \le 2$, 
indicated by red and purple,  
are due to a local bath.     
For presentation purpose (to avoid a crowded set of lines) 
the red couplings that originate from $\nu \mathcal{L}^{(\nu)}$ and $\eta \mathcal{E} \mathcal{L}^{(\tilde{\mathcal{E}})}$
assume that the local bath is positioned at bond~$x'$,
while purple couplings that originate  from $\eta c \mathcal{L}^{(\tilde{c})}$
assume that the bath is positioned at bond~$x''$. }
\end{figure}

A schematic representation of $\rho_x(r)$ and the couplings 
is given in \Fig{fig:schematic-transitions-all}.
The coherent hopping that is generated by $\mathcal{L}^{(c)}$ 
couples $\rho_x(r)$ to $\rho_x({r{\pm}1})$ and to $\rho_{x{+}1}(r{\pm}1)$, 
while $\mathcal{L}^{(\mathcal{E})}$ contribute ``on-site'' potential.
The noise operator $\nu\mathcal{L}^{(\nu)}$ include the Pauli-terms that were discussed previously,     
and an additional term that couples the $r={\pm}1$ elements.
Together with the friction operator $\eta \mathcal{E} \mathcal{L}^{(\tilde{\mathcal{E}})}$,
the Pauli terms induce the asymmetric $x{\pm}1$ stochastic transitions of \Eq{e5} along $r{=}0$.
The second friction term $\mathcal{L}^{(\bar{c})}$  
consists of non-Pauli terms that allow extra $\mathcal{L}^{(c)}$-type couplings,   
and in particular extra $x{\pm}2$ transitions within the strip $|r|=0,1,2$.

\sect{The spectrum for a non-disordered ring}
For a non-disordered ring the super-matrix $\mathcal{L}$ is invariant under $x$-translations, 
and therefore we can switch to a Fourier basis 
where the representation is $\rho(r;q)$. 
Due to Bloch theorem, the matrix decompose into $q$-blocks in this basis.  
Thus in order to find the eigenvalues $\lambda_{q,s}$ and the corresponding eigenmodes 
we merely have to handle a one dimensional tight binding $\ket{r}$~lattice.  
See Methods. 
A representative spectrum is provided in \Fig{f2}a. 
Consider first the $q{=}0$ eigenstates.
For $q{=}0$ the $c$-dependent couplings are zero.
For infinite temperature ($\eta{=}0$) the only non-zero coupling 
is between $\ket{r{=}\pm1}$ due to a non-Pauli term in \Eq{eLnu}.
Consequently the $q{=}0$ block contains the NESS $\ket{r{=}0}$ 
(which is merely the identity matrix in the standard basis),
along with a pair of non-trivial decoherence modes $\ket{\pm}$,  
and a set of uncoupled decoherence modes $\ket{r= \pm 2, \pm 3,...}$. 
The corresponding $\lambda_{q,s}$ eigenvalues (for $\eta{=}0$) are: 
\beq 
\lambda_{0,0}  &=& 0 \ \text{(NESS)} \\
\lambda_{0,\pm}  &=&  2\nu \pm \sqrt{\nu^2 - \mathcal{E}^2}  \\
\lambda_{0,s}  &=& 2\nu + i\mathcal{E} s, \ \ (s = \pm 2, \pm 3,...)  \label{eLadder}
\eeq
The $\ket{\pm}$ modes become over-damped for small bias, 
while the $|s|>1$ decoherence modes are always under-damped. 
%
Considering the $q$ dependence of the eigenvalues $\lambda_{q,s}$ we get several bands, 
as illustrated in \Fig{f2}a.   
Our interest below is in the relaxation modes that are associated with $\lambda_{q,0}$,  
and determine the long time spreading.

\sect{The NESS}
At finite temperature ($\eta>0$) there are extra couplings that lead to a modified NESS.
In leading order the NESS eigenstate is $\ket{0} + \alpha_0 \ket{1} + \alpha_0^{*} \ket{-1}$ with
\beq
\alpha_0 \ \ = \ \ \dfrac{3\nu - i\mathcal{E}}{3 \nu^2 + \mathcal{E}^2} \, \eta c 
\eeq
Reverting back to the standard representation we get
\beq \label{eq:rho-steady-state-maintext}
\rho^{(\text{NESS})} 
\ = \ \dfrac{1}{L}\left( \id + \alpha_0 e^{+i\bm{p}} + \alpha_0^{*} e^{-i\bm{p}}\right)
\eeq
From this we can deduce the steady state momentum distribution \cite{SM5}, 
namely, ${ p(k) \equiv  \BraKet{k}{\rho^{(\text{NESS})}}{k}}$.
The result in leading order is 
\beq \label{pk}
p(k) \propto  \exp\left[\dfrac{2 \eta c}{3 \nu^2 + \mathcal{E}^2} \left( 3 \nu \cos{(k)} +  \mathcal{E} \sin{(k)}\right)\right] 
\eeq
For $\mathcal{E}=0$ this expression is consistent 
with the canonical expectation $\exp(-\beta \bm{H}^{(c)})$.

\sect{The current}
For non-zero field ($\mathcal{E} \ne 0$) the NESS momentum distribution is shifted.
The expression for the current operator is complicated \cite{SM2}, 
but the net NESS current comes out a simple sum of stochastic and coherent terms:    
\beq
\label{eINESS}
I_x &=& \frac{1}{L} \left(  (w^+_x - w^-_x) - c\, \im(\alpha_0)  \right) 
\\ \label{eq:I} 
&=& \frac{1}{L}  \left[ 1 + \frac{c^2 }{6\nu^2 + 2\mathcal{E}^2} \right] 2\eta \mathcal{E} 
\ \equiv \ \frac{1}{L} v
\eeq
We shall further illuminate the physical significance of the second term below. 
In contrast with the stochastic case, the drift current 
might be non-monotonic in $\mathcal{E}$, see \Fig{f1}a.
Furthermore, there is a convex range where the second
derivative of $I(\mathcal{E})$ is positive.
 
The {\em convexity} of the current in some $\mathcal{E}$ range,
implies a counter intuitive effect: current may become larger due to disorder.
The argument goes as follows: Assume that the sample is divided into two regions, 
such that $\mathcal{E}_x$ is constant in each region, 
but slightly smaller (larger) than $\mathcal{E}$ in the first (second) region.
Due to the convex property it is implied that the current will be larger.
Extending this argument for a general non-homogeneous 
(i.e. disordered) field, with the same average bias $\mathcal{E}$,  
we expect to observe a {\em larger} NESS current.
This is indeed confirmed in \Fig{f1}a, while additional 
examples are given and discussed quantitatively in \cite{SM6}.

\sect{The Diffusion}
An optional way to derive \Eq{eq:I} is to expand $\lambda_{q,0}$ in $q$, 
to obtain~$v$. The second order term gives the diffusion coefficient. Namely, 
\beq \label{eVD} 
\lambda_{q,0} \ \ = \ \ \ i v q + D q^2 + O(q^3) 
\eeq
It is therefore enough to determine $\lambda_{q,0}$ via second order perturbation theory 
with respect to the $q{=}0$ eigenstates. To leading order in $\eta$, a lengthy calculation leads to a result 
that is consistent with the Einstein relation, namely 
\beq \label{eq:einstein-relation}
\frac{v}{D} = \frac{\mathcal{E}}{T}, 
\ \ \ \ \ \ \mbox{[valid in leading order]}
\eeq  
Thus, to leading order, $D$ is given by \Eq{eDP}, 
multiplied by the expression in the square brackets in \Eq{eq:I}. 
We see that with coherent transitions, for zero bias, 
this expression takes the form  ${D=\nu + D_{\ell} }$, 
where ${D_{\ell} = c^2/(6\nu)}$. 
The latter can be interpreted as a Drude-type result ${D_{\ell} =  \ell^2/\tau}$,
with relaxation time ${\tau \sim 1/\nu}$ 
and mean free path ${\ell \sim c\tau}$.
A similar expression has been obtained in \cite{MadhukarPost1977,EspositoGaspard2005} 
for a chain with noisy sites.
In the other extreme of large bias \Eq{eDP} implies 
that ${ D_{\ell} = (1/2) |c/\mathcal{E}|^2 \nu }$. 
This result, like the Drude result, can be regarded 
as coming from FGR transitions. But now the transitions 
are between Bloch site-localized states.  
Namely, we have hopping between neighboring sites (${ \ell \sim 1 }$), 
with rate of the transitions ($1/\tau$) that is suppressed  
by a factor $|c/\mathcal{E}|^2$. The suppression factor reflects 
the first-order-perturbation-theory overlap of Bloch-localized wavefunctions. 
To summarize: we can say that $D_{\ell}$ exhibits a crossover 
from Drude-type transport to hopping-type transport as the 
field $\mathcal{E}$ is increased.

In fact we can proceed beyond leading order, 
and calculate $D$ up to second order in $\eta$, see \cite{SM5}.
Here we cite only the zero bias result: 
\beq  \label{eD}
D \ = \  \left[ 1  +  \dfrac{c^{2}}{6 \nu^2} - \dfrac{c^{2}}{4 T^2} \right] \nu 
\eeq
In view of the Drude picture this result is surprising. Namely, 
one would expect ${D_{\ell} \sim c^2 \tau}$ to be replaced 
by ${D_{\ell} \sim \avg{v^2} \tau}$, and hence one would expect 
the replacement ${c^2 \mapsto (1 - [1/8](c/T)^2) c^2}$ 
due to the narrowing of the momentum distribution, see Methods. 
However the current result indicates that the leading correction 
is related to a different mechanism. Indeed, using a semiclassical perspective,
the coupling to the bath involves a $\cos{(p)}$ factor, see Methods. 
The zero order diffusion with rate $\nu$ arises due to stochastic term 
in the equation of motion for $\dot{x}$ that involves a $\sin{(p)}$ factor,   
Consequently, due to thermal averaging, ${\nu \mapsto (1 - [1/8](c/T)^2) \nu}$, 
which explains, up to a factor of~2, the third term in \Eq{eD}.
We have repeated this calculation also for a Caldeira-Leggett dissipator, 
and also for an $\mathcal{L}^{(\text{S})}$ dissipator. 
For the former the expected $(c/T)$ correction to $D_{\ell}$ 
appears, but has a different numerical factor, 
while for the latter the correction comes out with an opposite sign.
We can show analytically that the discrepancies 
are due to the modification of the correlation time \cite{qssXs-prep}.

\sect{Disordered ring}
The so called stochastic field $\mathcal{E}_x/T$ is responsible for the asymmetry of the incoherent transitions. 
Following Sinai we assume that it has a random component that is (say) box-distributed.  
From the works of Sinai and Derrida \cite{Sinai1983,DerridaPomeau1982,Derrida1983} we expect a sliding transition as $\mathcal{E}/T$ 
exceeds a critical value of order $(\sigma_{\mathcal{E}}/T )^2$. 
Strongly related is the delocalization transition 
\cite{Hatano1996,Hatano1997,Hatano1998,Amir2015,Amir2016,HurowitzCohen2016}
for which the critical value is smaller by a numerical factor. 
Disregarding this factor we expect 
\beq
\mathcal{E}_c \ \ \approx  \ \  \frac{1}{T}\sigma_{\mathcal{E}}^2
\eeq
In the purely stochastic model, for $\mathcal{E} > \mathcal{E}_c$ the relaxation is expected 
to be under-damped due to a delocalization transition that leads to the appearance 
of complex eigenvalues at the vicinity of $\lambda=0$.
          
The question arises how this transition is affected by quantum coherent hopping. 
The naive expectation would be to witness a smaller tendency for localization 
in the relaxation-spectrum because we add coherent bypass that enhances the transport. 
But surprisingly the numerical results of \Fig{f2}b show that the effect goes in the opposite direction:  
for non-zero~$c$, some eigenvalues become real, indicating stronger effective disorder.

\sect{Enhanced effective disorder}
We turn to provide an explanation for observing enhanced effective disorder due to coherent hopping. 
On the basis of the non-disordered ring analysis, the relaxation modes occupy mostly the ${|r|=0,1}$ diagonals 
of $\rho$, and therefore it makes sense to exclude couplings to the higher diagonals.
We verify  that this does not change the qualitative picture in \Fig{f2}b (gray vs green symbols). 
The effect of the ${|r| = 1}$ band is to introduce virtual coherent transitions between diagonal 
elements \cite{SM7}. Hence we end up with an effective single-band stochastic equation with transition rates  
\beq \label{eq:eff}
w^{\pm}_x \ &=& \ \nu + \nu_x \pm \eta \mathcal{E}_x \ \equiv \ w_x \exp(\pm \tilde{\mathcal{E}_x})\\
\nu_x \ &=& \dfrac{c^{2}}{2} \dfrac{\nu - \lambda}{(2 \nu - \lambda)^2 + \mathcal{E}^2_x - \nu^2}
\eeq
The disorder that is associated with $\nu_x$ is hermitian, namely, 
it does not spoil the symmetry of the transitions, it merely implies that 
the we have a tight binding model with random couplings  
that have some dispersion ${\sigma^2_{\perp} \equiv  \var (w_x) \propto c^4}$.
This is known as {\em resistor network} (RN) disorder.
In contrast, the $\pm \eta \mathcal{E}_x$ term induces {\em asymmetric} transitions.
This type of non-hermitian Sinai-type disorder is characterized by the dispersion 
${\sigma^2_{\parallel} \equiv  \var(\tilde{\mathcal{E}}_x)}$.
The latter translates after non-hermitian gauge transformation 
to (hermitian) diagonal disorder, with ill-defined boundary conditions.
The procedure to handle both types of disorder has been discussed in \cite{HurowitzCohen2016} following \cite{Hatano1997}. 
One defines an hermitian RN matrix by setting $\tilde{\mathcal{E}}_x{=}0$ in \Eq{eq:eff}.  
The RN matrix has a real spectrum with eigenstates that are characterized by inverse localization length that is dominated 
by the RN-disorder, namely, ${\kappa(\lambda) \propto \sigma_{\perp}^2 \lambda}$.  
Adding back the field $\mathcal{E}_x$, the eigenstates remain localized (with real eigenvalues) 
only in regions where localization is strong enough, that is, ${\kappa(\lambda) > \mathcal{E}/T}$. 
Estimating  ${\sigma_{\perp}}$, see \cite{SM7} for an explicit expression, 
we deduce that the additional RN disorder is responsible for the observed numerical result.

\sect{The Wannier-Stark Ladder}
{We shift our attention to the full Lindblad spectrum. 
In the absence of coupling to the bath, the eigenstates 
of the Hamiltonian \Eq{e02} are Bloch localized.
Each eigenstate occupies a spatial region ${\sim}c/\mathcal{E}$, 
and the corresponding eigen-energies form a ladder 
with spacing $\mathcal{E}$, that reflects the frequency 
of the Bloch oscillations. Weak coupling to the bath 
leads to damping of the Bloch oscillations. 
This is reflected by the Lindblad spectrum.
For the ${q=0}$ modes we have obtained \Eq{eLadder}, 
where we see that the eigenvalues acquire a real part, 
but maintain the ladder structure.  
But for non-zero $q$ the $\mathcal{L}^{(c)}$ term couples 
the modes to the perturbation that is created at the ${|r|<2}$ region  
by $\mathcal{L}^{(\nu)}$, see \Eq{eLc} and \Eq{eLnu} of the Methods. 
This results in a deformation of the ladder.   
Namely, the ladder consists of bands, 
and the number of bands that are deformed equals the Bloch localization length. See \Fig{fLadder}.

\begin{figure}
\includegraphics[]{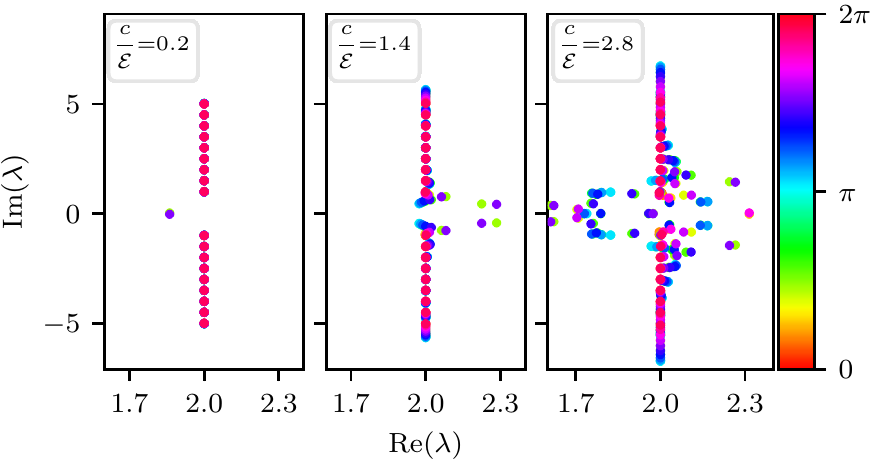}                                                      %
\caption{\label{fLadder}
{\bf Blurring of the Wannier-Stark Ladder.}
Here the parameters are $\nu=1, \eta=0.03, \mathcal{E}=0.5$ and $L=21$. 
The parameter ${c/\mathcal{E}}$ 
determines the localization range of the Bloch eigenstates.
The small $q$ modes are weakly coupled by $\nu$ 
and therefore maintain the $\mathcal{E}$ spacing 
as implied by \Eq{eLadder}. }
\end{figure}

\sect{Regime diagram}
We would like to place our results in the context 
of the vast quantum dissipation literature. The prototype model of
Quantum Brownian Motion (QBM), aka the Caldeira-Leggett model, 
involves coupling to a single bath that exerts a fluctuating homogeneous field of force. 
In the classical framework it leads to the standard Langevin equation
\beq  \label{eLNGV} 
\mass \ddot{x} \ = \ \mathcal{E} - \eta \dot{x} + f(t)  
\eeq
and \Eq{e1} becomes the standard Fokker Planck equation. 
In the tight-binding framework we have the 
identification ${ \mass \mapsto 1/(c a^2) }$, 
where $a$ is the lattice constant.
The standard QBM model features a single dimensionless parameter, 
the scaled inverse temperature $\beta$, which is the ratio between 
the thermal time $1/T$ and damping time ${\mass/\eta}$. 
In the lattice problem we can define two dimensionless parameters
\beq
\alpha &=& \frac{1}{2\pi} \eta a^2 \ \ \ \ \ \ \ \ \ \ \mbox{scaled friction} \\
\theta &=& \frac{T}{c} \ = \ \frac{\nu}{2\eta c}  \ \ \ \ \ \ \mbox{scaled temperature}
\eeq
Accordingly $\beta = \alpha/\theta$. 
Note that in our model we set the units such that ${a=1}$, 
hence, disregarding $2\pi$ factor, 
our scaled friction parameter $\eta$ is the same as $\alpha$.

\begin{figure}
\includegraphics[]{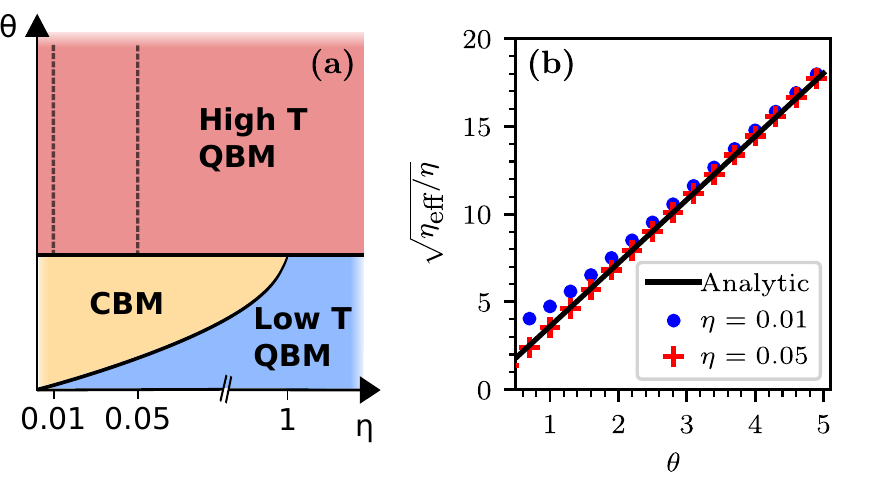}
\caption{\label{fRegimes}
{\bf The Brownian Motion regime diagram.}
(a) The various regions in the  $(\eta,\theta)$ diagram are indicated. 
We distinguish between the Classical-like Brownian Motion (CBM) region; 
the low-temperature QBM region where memory effects dominates;
and the high-temperature QBM region that has been discussed in this article.
The Lindblad correction to the Ohmic master equation is negligible above the solid ${\theta\sim 1}$ line. 
(b) The effective friction coefficient $\eta_{\text{eff}}$ of \Eq{eq:v-eta-eff} 
is determined numerically along the two dashed lines of panel~(a), 
and compared with the analytical prediction of \Eq{eq:I}.   
The parameters are $\nu=1, \mathcal{E}=0.5$ and $L=500$. 
For lower $\theta$, the Lindblad correction becomes important (not shown).}
\end{figure}

The standard analysis of QBM \cite{HakimAmbegaokar1985} reveals 
that quantum-implied memory effects 
are expressed in the regime ${\beta \gg 1 }$, 
where a transient $\log(t)$ spreading is observed
in the absence of bias, followed by diffusion.  
Most of the quantum dissipation literature, 
regarding the two-site spin-boson model \cite{LeggettEtAlDynamicsTwoLevel1987} 
and regarding multi-site chains \cite{aslangul1986quantum,AslangulPeriodicPotential1987},  
is focused in this low temperature regime,
where significant deviations from the classical
predictions are observed for large $\alpha$ of order unity.   
In contrast, our interest is in the ${\alpha, \beta \ll 1 }$ regime.

Our $(\eta,\theta)$ regime diagram \Fig{fRegimes} is roughly divided 
into two regions by the line ${\theta \sim 1}$. 
Along this line the thermal de-Broglie wavelength 
of the particle is of order of the lattice constant, 
hence it bears formal analogy to the analysis
of QBM in cosine potential \cite{Fisher1985QuantumBrownianPeriodic}, where it marks the border 
to the regime where activation mechanism comes into action. 
In our tight binding model we have a single band, 
hence transport via thermal activation is not possible. 
Rather, in the ${\theta > 1}$ regime, where $T\gg c$, 
the momentum distribution within the band is roughly flat,
and the drift is dictated by \Eq{eq:I}, that is,
\beq \label{eq:v-eta-eff}
v \ = \ 2\eta\mathcal{E} + \frac{1}{\eta_{\text{eff}}} \mathcal{E}
\eeq
where ${\eta_{\text{eff}} \approx 12 \theta^2 \eta \,}$ for weak field.  
The  low temperature regime  $\theta \ll1$  has not been addressed in this work, because the Ohmic master equation fails to reproduce canonical equilibration in this regime  (see Methods). Still we would like to illuminate what we get (not presented) if this aspect is corrected. 
In the regime ${\theta \ll 1}$ the momentum distribution becomes 
much narrower (only low energy momenta are populated) 
and therefore ${\eta_{\text{eff}} \sim \eta}$ as implied by \Eq{eLNGV}.  
This we call Classical-like Brownian motion (CBM) regime. 
Once the coupling to the bath is not the simple $x$-coupling 
of the Caldeira-Leggett model, a numerical prefactor is expected 
(see Methods for detailed argument).

\vspace*{5mm}
\noindent {\Large\bf\textsl{Discussion}} 
\vspace*{1mm}

%
There is a rich literature regarding Quantum Brownian motion (see \cite{Schwinger1961,HakimAmbegaokar1985,GRABERT1988115,HanggiQBM2005} and references within).
In the condensed matter literature it is common to refer 
to the Caldeira-Leggett model \cite{Caldeira1983,CALDEIRA1983374}, 
where the particle is linearly coupled to a bath of harmonic oscillators 
that mimic an Ohmic environment. 
Some works study the motion of a particle in a periodic potential,   
possibly with bias, aka washboard potential \cite{Schmid1983,Fisher1985QuantumBrownianPeriodic,Fisher1985QuantumBrownianPeriodic,AslangulPeriodicPotential1987}, 
while other refer to tight binding models \cite{Weiss1985,aslangul1986quantum,Weiss1991} as in this work.  
The focus in those papers is mostly on non-Markovian effects: at low temperatures 
the fluctuations are not like ``white noise", and are dominated by a high frequency cutoff $\omega_c$.
Consequently the handling of long-time correlations becomes tricky. 
In this context the low temperature dependence of the diffusion and the mobility is modified for ${\alpha>1/2}$ and ${\alpha>1}$.

The line of study in the above models has assumed that the fluctuations that are induced 
by the bath are uniform in space. In some other works the dynamics of a particle that interacts with local baths 
has been considered. In such models the fluctuations acquire finite correlation length in space  
\cite{Cohen1997,MadhukarPost1977,Kumar1985,EspositoGaspard2005,Dibyendu2008,Amir2009,lloyd2011quantum,Moix_2013,CaoSilbeyWu2013,Moix_2013,Kaplan2017ExitSite,Kaplan2017B}.
The extreme case, as in this work, are tight binding models where the coupling is to uncorrelated baths that seat on different sites or bonds.
Studies in this context assume bath that are connected at the end points \cite{znidaricOpenQuantumChain2010}, 
or baths that act as noise source \cite{MadhukarPost1977}. 
Ref. \cite{EspositoGaspard2005} has analyzed the spectral properties for a chain with noisy sites,  
Ref. \cite{Amir2009} has considered colored noise sources strongly coupled to each site, 
Ref. \cite{EislerCrossoverBallistiDiffusiveQuantum_2011} has considered noisy transitions on top of coherent transitions,
Ref. \cite{znidaricDisorderDephasing2013} has considered transport in the presence of dephasing and disorder,
Ref. \cite{Moix_2013} has considered numerically transport properties of a noisy system with static disorder, 
and \cite{Han_local_bath_diffusion_2018} has addressed some bounds in the absence of disorder.
The basic question of transport in a tight-binding models has resurfaced 
in the context of excitation transport in photosynthetic light-harvesting complexes    \cite{amerongen2000photosynthetic,ritz2002quantum,plenio2008dephasing,FlemingCheng2009,Rebentrost_2009,Alan2009,Sarovar_2013,higgins2014superabsorption,celardo2012superradiance,park2016enhanced}.

It is quite surprising that all of the above cited works have somehow avoided the confrontation 
with themes that are familiar from the study of stochastic motion in random environment.
Specifically we refer here to the extensive work by Sinai, Derrida, and followers \cite{Dyson1953,Sinai1983,DerridaPomeau1982,Derrida1983,havlin1987diffusion,Bouchaud1990,BOUCHAUD1990a}, 
and the studies of stochastic relaxation \cite{HurowitzCohen2016,HurowitzCohen2016a}  which is related to the works of Hatano, Nelson and followers \cite{Hatano1996,Hatano1997,Hatano1998,Shnerb99,LubenskyNelson2000,LubenskyNelson2002,Amir2015,Amir2016,HurowitzCohen2016,HurowitzCohen2016a}.   
Clearly we have here a gap that should be bridged.

In this article we have studied transport properties along a chain 
taking into account several themes 
that have not been combined in past studies: 
{\bf (a)}~ The baths on different bonds are not correlated in space;
{\bf (b)}~ The baths are not just noise - the temperature is high but finite; 
{\bf (c)}~ Without coherent hopping it is the Sinai-Derrida-Hatano-Nelson model which exhibits sliding and delocalization transitions;
{\bf (d)}~ Without baths it is a disordered chain with Anderson localization;
{\bf (e)}~ The bias might be large such that Bloch dynamics is reflected.

The ``small'' parameter in our analysis is the inverse temperature.
The following observations have been highlighted: 
{\bf (1)}~ The NESS current is the sum of stochastic and quasi-coherent terms; 
{\bf (2)}~ It displays non-monotonic dependence on the bias, as shown in \Fig{f1}, due to crossover from Drude-type to hopping-type transport;  
{\bf (3)}~ Disorder may increase the current due to convex property; 
{\bf (4)}~ The interplay of stochastic and coherent transition is reflected in the Lindblad spectrum;  
{\bf (5)}~ In the presence of disorder the quasi-coherent transitions enhance the localization of the relaxation modes. 
Thus, with regard to the Sinai-Derrida sliding transition, and the strongly related Hatano-Nelson delocalization transition, 
we find that adding coherent transitions ``in parallel" have in some sense opposite effects: 
on the one hand they add bypass for the current (point (1) above), 
but on the other hand they enhance the tendency towards localization (point (5) above).
Some of our results might be relevant to studies of optimal transport efficiency \cite{R1,R2} and the quantum Goldilocks effect \cite{R3}.

\vspace*{5mm}
\noindent {\Large\bf\textsl{Methods}} 
\vspace*{1mm}

\sect{Master equation for disordered chain}
A pedagogical presentation of the procedure for the construction of an Ohmic master
equation for a two site system, and then for a chain, is presented in \cite{SM1}. 
Each bond has a different bath, and therefore can experience different temperature and friction.
Accordingly we can have disorder that originates either from the Hamiltonian 
(say random $\mathcal{E}_x$ as assumed in the main text), 
or from having different baths (random  $\mathcal{\nu}_x$ or random $\eta_x$). 
This extra disorder can be incorporated in a straightforward manner, 
and does not affect the big picture.       
\\

\sect{Thermalization}
The Ohmic master equation for a Brownian particle, 
if the coupling to the bath were $-\bm{x}f(t)$,  
is the standard Fokker-Plank equation.
It leads to canonical thermal state for any friction and for any temperature. This is not the 
case for a discrete two level system. The agreement with the canonical result is guaranteed only 
to first order in $\eta$. This is reflected in \Eq{e5}. The same applies for a chain. 
Note that in this sense the Ohmic approximation is very different from the secular or Pauli approximation \cite{Breuer2002}, 
or specially constructed Davies Liovillian \cite{davies1976quantum}, 
that guarantee canonical thermalization.   
%

It is important to identify the ``small parameter" that controls
the deviation from canonical thermalization. The standard coupling via $\bm{x}$
induce transitions between neighboring momenta, and therefore the small parameter
is $\Delta/T$, where the level spacing $\Delta$ goes to zero in the $L\rightarrow\infty$ limit.   
But for local baths, the coupling is to $\delta(\bm{x}-x_{\alpha})$ scatterers,
that create transitions to all the levels within the band.
Therefore the small parameter in the absence of bias is $c/T$.
This assertion is confirmed numerically by inspection of the equilibrium
momentum distribution $p(k)$, see \cite{SM5}. 
We conclude that in the regime of interest ($\theta{>}1$)
the Ohmic master equation can be trusted, while for lower temperatures
we have to ``correct" it.
For the two site system the ``corrected" equation is the Bloch equation,
where the ratio of the rates is in agreement with detailed balance
(not just in leading order in~$\eta$). For a chain, we cannot justify the secular approximation,
and therefore the correction procedure is ill defined.
\\

\sect{The friction coefficient}
In the Caldeira-Leggett model for Brownian particle,
with interaction term $-{\bm{x}}f(t)$,
the bath induced fluctuations $f(t)$
are determined by a coupling constant $\eta$,
and by the bath temperature $T$,
such that at high temperatures the intensity
of the fluctuations is ${\nu=2\eta T}$.
The $\eta$ parameter is defined such that the
friction coefficient in \Eq{eLNGV} equals~$\eta$.
A straightforward generalizations \cite{Cohen1997}
shows that for interaction with local baths
$\sum_{\alpha} u_{\alpha}({\bm{x}}) f_{\alpha}(t)$,
with $u_{\alpha}(x)= u(x{-}x_{\alpha})$,   
the effective friction coefficient is
${\eta_{\text{eff}}(x) = \eta \sum_{\alpha} [u'_{\alpha}(x)]^2}$.
For homogeneous distribution of local baths that have
the same $\eta$, the friction coefficient becomes $x$-independent. 
In the model under consideration the coupling
to the baths is $\sum_{\alpha} \bm{W}_{\alpha} f_{\alpha}(t)$, 
where $\alpha$ labels the sites. 
Disregarding commutation, it can be written as
$2 \cos(\bm{p}) \sum_{\alpha} u_{\alpha}(\bm{x}) f_{\alpha}(t)$,
where the $u$-s are site localized. 
It follow that the effective friction coefficient is
momentum dependent, namely
${\eta_{\text{eff}} \sim  |\cos(p)|^2 \eta}$.
But for equilibration in the ${\theta < 1}$ regime
only low momenta are important
hence we expect, up to numerical prefactor
to observe ${\eta_{\text{eff}} \approx \eta }$.
The failure to observe this result is due to improper
thermalization, as discussed in a previous paragraph.
\\

\sect{Positivity}
Irrespective of~$\eta$, there is another complication with the Ohmic master equation. 
If the temperature is low (small $\nu$) the relaxation may lead to a sub-minimal wavepacket 
that violates the uncertainty principle. This reflects the observation that 
the Ohmic master equation is not of Lindblad form, and violates the {\em positivity} requirement.
The minimal correction required to restore positivity is to couple $\bm{V}$ 
to an extra noise source of intensity 
\beq
\nu_{\eta} \ \ = \ \ \frac{\nu}{(4T)^2} \ \ = \ \ \frac{\eta^2}{4\nu}
\eeq  
This term is essential in the low temperature regime.
We have verified numerically that the extra noise term can be neglected 
in the high temperature regime where our interest is focused.
\\

\sect{On-site dissipators}
The model of \cite{EspositoGaspard2005} combines Hamiltonian term 
with dissipator of the $\mathcal{L}^{(\text{S})}$ type
that originates from couplings via  ${\bm{W}_x := \bm{Q}_x }$, 
where ${\bm{Q}_x = \ket{x}\bra{x}}$.    
Such dissipator leads to off-diagonal dephasing
that is generated by $\bm{Q}_x \rho \bm{Q}_x$ terms, 
and therefore excludes the possibility for inter-site stochastic transitions. 
Similar remark applies to the familiar Caldeira-Leggett model 
of Quantum Brownian motion \cite{Caldeira1983}, where the coupling is via  ${\bm{W} := \bm{x} }$.   
Namely, it is a single bath that exerts a fluctuating homogeneous force 
that affects equally all the sites, as in \cite{aslangul1986quantum}. 
In our model the dissipation effect is local: many uncorrelated local baths.
\\

\sect{Pauli dissipator}
A conventional Pauli-type dissipator is obtained 
if we drop some of the terms in the Ohmic dissipator of \Eq{e2}. 
Namely, 
\beq \nonumber
&& \mathcal{L}^{(\text{Pauli})} \rho \ =  \ -(w^{+}+w^{-})\rho  
\\ \nonumber
&& \ + \sum_{x} \left( w^{+}  \bm{D}_x\+ \rho \bm{D}_x \ + \  w^{-}  \bm{D}_x \rho \bm{D}_x\+ \right)
\\  \label{e3} 
&& - \gamma \left( \rho - \sum_{x} \bm{Q}_x  \rho \bm{Q}_x  \right)
\eeq
The transition rates between sites, $w^{\pm}=[\nu \pm \eta \mathcal{E}]$, 
are in agreement with FGR. 
For completeness, we added here a $\gamma$ term that represents 
optional off-diagonal decoherence due to on-site noise. 
\\

\sect{Ring configuration}
For numerical treatment, and for the purpose of studying relaxation dynamics, we close the chain into a ring.  
This means to impose periodic boundary conditions. 
With uniform field $\mathcal{E}$, one encounter a huge potential drop at the boundary.   
To avoid this complication we assume that the boundary bond has an infinite temperature, 
hence the formation of a stochastic barrier is avoided, 
and the circulation of the stochastic field (aka {\em affinity}) becomes $\mathcal{E}/T$ as desired. 
In the analytical treatment of a clean ring we assume 
that $\nu$, and $\eta$ and $\mathcal{E}$ in the master equation are all uniform, 
such that invariance under translation is regained. 
This cheat is valid for large ring if $\rho$ is banded, 
reflecting a finite spatial correlation scale.  
See numerical verification in \cite{SM2}. 
\\

\sect{The Bloch eigenstates of a clean ring}
The $\mathcal{L}$ super-operator in the Bloch $(r;q)$ basis, 
decomposes into $q$ Blocks. Each block can be written 
as a sum of terms, as in \Eq{e4}, that operate over 
a one dimensional tight binding $\ket{r}$ lattice. 
In this representation the coherent dynamics is generated by 
\beq
\mathcal{L}^{(\mathcal{E})} &=& -i\sum_r \ket{r}r\bra{r} 
\\ \label{eLc}
\mathcal{L}^{(c)} &=& \sin(q/2) [\mathcal{D}_{\perp}^{\dag} - \mathcal{D}_{\perp} ] 
\eeq
where $\mathcal{D}_{\perp}=\sum_r \ket{r{+}1}\bra{r}$. 
For the $\nu$ induced stochastic transitions we have 
\beq \label{eLnu}
\mathcal{L}^{(\nu)} =&& -2 \ + \ 2\cos(q) \kb{0}{0} \\
&& + \left(\kb{1}{{-1}} + \kb{{-1}}{1}  \right) 
\eeq
where the last term is non-Pauli.
The Pauli-type friction term 
takes the form:
\beq
\mathcal{L}^{(\tilde{\mathcal{E}})} = -2i \sin{(q)} \kb{0}{0}
\eeq
And the additional friction terms are:
\beq  
\mathcal{L}^{(\tilde{c})} &=&  \frac{1}{2}\cos{(q/2)} [\mathcal{D}_{\perp}+\mathcal{D}_{\perp}^{\dag}]  \\
&& + \dfrac{1}{2} \cos(3q/2) \left[  \kb{\pm 1}{0} - \kb{0}{\pm 1} \right] \\
&& + \dfrac{1}{2}\cos(q/2) \left[ \kb{{\mp 2}}{\pm 1} - \kb{\pm 1}{{\mp 2}} \right] 
\eeq
Note that the zero eigenvalue belongs to the $q=0$ block.
Some more details are provided in \cite{SM5}. 
\\

\sect{Diffusion at finite temperature}
The Drude type term in the expression for the diffusion \Eq{eD} 
is up to numerical prefactor $\braket{v_k^2}\tau$, 
where $\braket{v_k^2} = c^2/2$ for uniform momentum distribution. 
At finite temperature this distribution \Eq{pk} is not uniform.
Here we consider zero bias and get
\beq
\braket{v_k^2}  = 
\int_{-\pi}^{\pi} [c \sin{(k)}]^2 p(k) dk 
\approx  \left[ 1 - \dfrac{1}{8} (\beta c)^2 \right] \dfrac{c^{2}}{2}
\ \ \ \ \ \ \ 
\eeq
where $\beta=1/T$, and recall that $\eta=\nu/(2T)$.
The analytical calculation in \cite{SM5} leads to a different result 
which implies that the expression in the square brackets      
should be replaced by $[1-6 \eta^2]$, which means that the 
relevant dimensionless parameter is $\nu/T$ and not $c/T$.    
\Fig{fig:rho-p-canonical} of \cite{SM5}b confirms this statement.

\clearpage
\onecolumngrid

\sect{Acknowledgment} 
This research was supported by the Israel Science Foundation (Grant  No.283/18).
\\

\hide{

\sect{Contributions}
Both authors have contributed to this article. D.S has carried out the analysis, including numerics and figure preparation. The themes of the study and the text of Ms have been discussed, written and iterated jointly by D.C. and D.S.
\\

\sect{Competing interests}
The authors declare no competing financial interests.
\\

\sect{Corresponding authors} 
Correspondence to D.S. [dekels@post.bgu.ac.il] and/or D.C.  [dcohen@bgu.ac.il]

}




\begin{thebibliography}{63}%
\makeatletter
\providecommand \@ifxundefined [1]{%
 \@ifx{#1\undefined}
}%
\providecommand \@ifnum [1]{%
 \ifnum #1\expandafter \@firstoftwo
 \else \expandafter \@secondoftwo
 \fi
}%
\providecommand \@ifx [1]{%
 \ifx #1\expandafter \@firstoftwo
 \else \expandafter \@secondoftwo
 \fi
}%
\providecommand \natexlab [1]{#1}%
\providecommand \enquote  [1]{``#1''}%
\providecommand \bibnamefont  [1]{#1}%
\providecommand \bibfnamefont [1]{#1}%
\providecommand \citenamefont [1]{#1}%
\providecommand \href@noop [0]{\@secondoftwo}%
\providecommand \href [0]{\begingroup \@sanitize@url \@href}%
\providecommand \@href[1]{\@@startlink{#1}\@@href}%
\providecommand \@@href[1]{\endgroup#1\@@endlink}%
\providecommand \@sanitize@url [0]{\catcode `\\12\catcode `\$12\catcode
  `\&12\catcode `\#12\catcode `\^12\catcode `\_12\catcode `\%12\relax}%
\providecommand \@@startlink[1]{}%
\providecommand \@@endlink[0]{}%
\providecommand \url  [0]{\begingroup\@sanitize@url \@url }%
\providecommand \@url [1]{\endgroup\@href {#1}{\urlprefix }}%
\providecommand \urlprefix  [0]{URL }%
\providecommand \Eprint [0]{\href }%
\providecommand \doibase [0]{http://dx.doi.org/}%
\providecommand \selectlanguage [0]{\@gobble}%
\providecommand \bibinfo  [0]{\@secondoftwo}%
\providecommand \bibfield  [0]{\@secondoftwo}%
\providecommand \translation [1]{[#1]}%
\providecommand \BibitemOpen [0]{}%
\providecommand \bibitemStop [0]{}%
\providecommand \bibitemNoStop [0]{.\EOS\space}%
\providecommand \EOS [0]{\spacefactor3000\relax}%
\providecommand \BibitemShut  [1]{\csname bibitem#1\endcsname}%
\let\auto@bib@innerbib\@empty
\bibitem [{\citenamefont {Hartmann}\ \emph {et~al.}(2004)\citenamefont
  {Hartmann}, \citenamefont {Keck}, \citenamefont {Korsch},\ and\ \citenamefont
  {Mossmann}}]{Hartmann_korsch_2004}%
  \BibitemOpen
  \bibfield  {author} {\bibinfo {author} {\bibfnamefont {T}~\bibnamefont
  {Hartmann}}, \bibinfo {author} {\bibfnamefont {F}~\bibnamefont {Keck}},
  \bibinfo {author} {\bibfnamefont {H~J}\ \bibnamefont {Korsch}}, \ and\
  \bibinfo {author} {\bibfnamefont {S}~\bibnamefont {Mossmann}},\ }\bibfield
  {title} {\enquote {\bibinfo {title} {Dynamics of bloch oscillations},}\
  }\href {\doibase 10.1088/1367-2630/6/1/002} {\bibfield  {journal} {\bibinfo
  {journal} {New Journal of Physics}\ }\textbf {\bibinfo {volume} {6}},\
  \bibinfo {pages} {2--2} (\bibinfo {year} {2004})}\BibitemShut {NoStop}%
\bibitem [{\citenamefont {Dubin}\ \emph {et~al.}(2006)\citenamefont {Dubin},
  \citenamefont {Melet}, \citenamefont {Barisien}, \citenamefont {Grousson},
  \citenamefont {Legrand}, \citenamefont {Schott},\ and\ \citenamefont
  {Voliotis}}]{dubin2006macroscopic}%
  \BibitemOpen
  \bibfield  {author} {\bibinfo {author} {\bibfnamefont {Francois}\
  \bibnamefont {Dubin}}, \bibinfo {author} {\bibfnamefont {Romain}\
  \bibnamefont {Melet}}, \bibinfo {author} {\bibfnamefont {Thierry}\
  \bibnamefont {Barisien}}, \bibinfo {author} {\bibfnamefont {Roger}\
  \bibnamefont {Grousson}}, \bibinfo {author} {\bibfnamefont {Laurent}\
  \bibnamefont {Legrand}}, \bibinfo {author} {\bibfnamefont {Michel}\
  \bibnamefont {Schott}}, \ and\ \bibinfo {author} {\bibfnamefont {Valia}\
  \bibnamefont {Voliotis}},\ }\bibfield  {title} {\enquote {\bibinfo {title}
  {Macroscopic coherence of a single exciton state in an organic quantum
  wire},}\ }\href {https://www.nature.com/articles/nphys196} {\bibfield
  {journal} {\bibinfo  {journal} {Nature Physics}\ }\textbf {\bibinfo {volume}
  {2}},\ \bibinfo {pages} {32} (\bibinfo {year} {2006})}\BibitemShut {NoStop}%
\bibitem [{\citenamefont {Nelson}\ \emph {et~al.}(2018)\citenamefont {Nelson},
  \citenamefont {Ondarse-Alvarez}, \citenamefont {Oldani}, \citenamefont
  {Rodriguez-Hernandez}, \citenamefont {Alfonso-Hernandez}, \citenamefont
  {Galindo}, \citenamefont {Kleiman}, \citenamefont {Fernandez-Alberti},
  \citenamefont {Roitberg},\ and\ \citenamefont
  {Tretiak}}]{nelson2018coherent}%
  \BibitemOpen
  \bibfield  {author} {\bibinfo {author} {\bibfnamefont {Tammie~R}\
  \bibnamefont {Nelson}}, \bibinfo {author} {\bibfnamefont {Dianelys}\
  \bibnamefont {Ondarse-Alvarez}}, \bibinfo {author} {\bibfnamefont {Nicolas}\
  \bibnamefont {Oldani}}, \bibinfo {author} {\bibfnamefont {Beatriz}\
  \bibnamefont {Rodriguez-Hernandez}}, \bibinfo {author} {\bibfnamefont
  {Laura}\ \bibnamefont {Alfonso-Hernandez}}, \bibinfo {author} {\bibfnamefont
  {Johan~F}\ \bibnamefont {Galindo}}, \bibinfo {author} {\bibfnamefont
  {Valeria~D}\ \bibnamefont {Kleiman}}, \bibinfo {author} {\bibfnamefont
  {Sebastian}\ \bibnamefont {Fernandez-Alberti}}, \bibinfo {author}
  {\bibfnamefont {Adrian~E}\ \bibnamefont {Roitberg}}, \ and\ \bibinfo {author}
  {\bibfnamefont {Sergei}\ \bibnamefont {Tretiak}},\ }\bibfield  {title}
  {\enquote {\bibinfo {title} {Coherent exciton-vibrational dynamics and energy
  transfer in conjugated organics},}\ }\href
  {https://www.nature.com/articles/s41467-018-04694-8} {\bibfield  {journal}
  {\bibinfo  {journal} {Nature communications}\ }\textbf {\bibinfo {volume}
  {9}},\ \bibinfo {pages} {2316} (\bibinfo {year} {2018})}\BibitemShut
  {NoStop}%
\bibitem [{\citenamefont {Dekorsy}\ \emph {et~al.}(2000)\citenamefont
  {Dekorsy}, \citenamefont {Bartels}, \citenamefont {Kurz}, \citenamefont
  {K{\"o}hler}, \citenamefont {Hey},\ and\ \citenamefont
  {Ploog}}]{dekorsy2000coupled}%
  \BibitemOpen
  \bibfield  {author} {\bibinfo {author} {\bibfnamefont {Thomas}\ \bibnamefont
  {Dekorsy}}, \bibinfo {author} {\bibfnamefont {Albrecht}\ \bibnamefont
  {Bartels}}, \bibinfo {author} {\bibfnamefont {Heinrich}\ \bibnamefont
  {Kurz}}, \bibinfo {author} {\bibfnamefont {Klaus}\ \bibnamefont
  {K{\"o}hler}}, \bibinfo {author} {\bibfnamefont {Rudolf}\ \bibnamefont
  {Hey}}, \ and\ \bibinfo {author} {\bibfnamefont {Klaus}\ \bibnamefont
  {Ploog}},\ }\bibfield  {title} {\enquote {\bibinfo {title} {Coupled
  bloch-phonon oscillations in semiconductor superlattices},}\ }\href@noop {}
  {\bibfield  {journal} {\bibinfo  {journal} {Physical review letters}\
  }\textbf {\bibinfo {volume} {85}},\ \bibinfo {pages} {1080} (\bibinfo {year}
  {2000})}\BibitemShut {NoStop}%
\bibitem [{\citenamefont {Madhukar}\ and\ \citenamefont
  {Post}(1977)}]{MadhukarPost1977}%
  \BibitemOpen
  \bibfield  {author} {\bibinfo {author} {\bibfnamefont {A}~\bibnamefont
  {Madhukar}}\ and\ \bibinfo {author} {\bibfnamefont {W}~\bibnamefont {Post}},\
  }\bibfield  {title} {\enquote {\bibinfo {title} {Exact solution for the
  diffusion of a particle in a medium with site diagonal and off-diagonal
  dynamic disorder},}\ }\href
  {http://journals.aps.org/prl/pdf/10.1103/PhysRevLett.39.1424} {\bibfield
  {journal} {\bibinfo  {journal} {Physical Review Letters}\ }\textbf {\bibinfo
  {volume} {39}},\ \bibinfo {pages} {1424} (\bibinfo {year}
  {1977})}\BibitemShut {NoStop}%
\bibitem [{\citenamefont {Weiss}\ and\ \citenamefont
  {Grabert}(1985)}]{Weiss1985}%
  \BibitemOpen
  \bibfield  {author} {\bibinfo {author} {\bibfnamefont {Ulrich}\ \bibnamefont
  {Weiss}}\ and\ \bibinfo {author} {\bibfnamefont {Hermann}\ \bibnamefont
  {Grabert}},\ }\bibfield  {title} {\enquote {\bibinfo {title} {Quantum
  diffusion of a particle in a periodic potential with ohmic dissipation},}\
  }\href {https://www.sciencedirect.com/science/article/pii/0375960185905171}
  {\bibfield  {journal} {\bibinfo  {journal} {Physics Letters A}\ }\textbf
  {\bibinfo {volume} {108}},\ \bibinfo {pages} {63--67} (\bibinfo {year}
  {1985})}\BibitemShut {NoStop}%
\bibitem [{\citenamefont {Kumar}\ and\ \citenamefont
  {Jayannavar}(1985)}]{Kumar1985}%
  \BibitemOpen
  \bibfield  {author} {\bibinfo {author} {\bibfnamefont {N.}~\bibnamefont
  {Kumar}}\ and\ \bibinfo {author} {\bibfnamefont {A.~M.}\ \bibnamefont
  {Jayannavar}},\ }\bibfield  {title} {\enquote {\bibinfo {title} {Quantum
  diffusion in thin disordered wires},}\ }\href {\doibase
  10.1103/PhysRevB.32.3345} {\bibfield  {journal} {\bibinfo  {journal} {Phys.
  Rev. B}\ }\textbf {\bibinfo {volume} {32}},\ \bibinfo {pages} {3345--3347}
  (\bibinfo {year} {1985})}\BibitemShut {NoStop}%
\bibitem [{\citenamefont {Roy}(2008)}]{Dibyendu2008}%
  \BibitemOpen
  \bibfield  {author} {\bibinfo {author} {\bibfnamefont {Dibyendu}\
  \bibnamefont {Roy}},\ }\bibfield  {title} {\enquote {\bibinfo {title}
  {Crossover from ballistic to diffusive thermal transport in quantum langevin
  dynamics study of a harmonic chain connected to self-consistent
  reservoirs},}\ }\href {\doibase 10.1103/PhysRevE.77.062102} {\bibfield
  {journal} {\bibinfo  {journal} {Physical review. E, Statistical, nonlinear,
  and soft matter physics}\ }\textbf {\bibinfo {volume} {77}},\ \bibinfo
  {pages} {062102} (\bibinfo {year} {2008})}\BibitemShut {NoStop}%
\bibitem [{\citenamefont {Amir}\ \emph {et~al.}(2009)\citenamefont {Amir},
  \citenamefont {Lahini},\ and\ \citenamefont {Perets}}]{Amir2009}%
  \BibitemOpen
  \bibfield  {author} {\bibinfo {author} {\bibfnamefont {Ariel}\ \bibnamefont
  {Amir}}, \bibinfo {author} {\bibfnamefont {Yoav}\ \bibnamefont {Lahini}}, \
  and\ \bibinfo {author} {\bibfnamefont {Hagai~B}\ \bibnamefont {Perets}},\
  }\bibfield  {title} {\enquote {\bibinfo {title} {Classical diffusion of a
  quantum particle in a noisy environment},}\ }\href
  {http://journals.aps.org/pre/abstract/10.1103/PhysRevE.79.050105} {\bibfield
  {journal} {\bibinfo  {journal} {Physical Review E}\ }\textbf {\bibinfo
  {volume} {79}},\ \bibinfo {pages} {050105} (\bibinfo {year}
  {2009})}\BibitemShut {NoStop}%
\bibitem [{\citenamefont {Lloyd}\ \emph {et~al.}(2011)\citenamefont {Lloyd},
  \citenamefont {Mohseni}, \citenamefont {Shabani},\ and\ \citenamefont
  {Rabitz}}]{lloyd2011quantum}%
  \BibitemOpen
  \bibfield  {author} {\bibinfo {author} {\bibfnamefont {Seth}\ \bibnamefont
  {Lloyd}}, \bibinfo {author} {\bibfnamefont {Masoud}\ \bibnamefont {Mohseni}},
  \bibinfo {author} {\bibfnamefont {Alireza}\ \bibnamefont {Shabani}}, \ and\
  \bibinfo {author} {\bibfnamefont {Herschel}\ \bibnamefont {Rabitz}},\
  }\bibfield  {title} {\enquote {\bibinfo {title} {The quantum goldilocks
  effect: on the convergence of timescales in quantum transport},}\ }\href
  {https://arxiv.org/abs/1111.4982} {\bibfield  {journal} {\bibinfo  {journal}
  {arXiv:1111.4982}\ } (\bibinfo {year} {2011})}\BibitemShut {NoStop}%
\bibitem [{\citenamefont {Moix}\ \emph {et~al.}(2013)\citenamefont {Moix},
  \citenamefont {Khasin},\ and\ \citenamefont {Cao}}]{Moix_2013}%
  \BibitemOpen
  \bibfield  {author} {\bibinfo {author} {\bibfnamefont {Jeremy~M}\
  \bibnamefont {Moix}}, \bibinfo {author} {\bibfnamefont {Michael}\
  \bibnamefont {Khasin}}, \ and\ \bibinfo {author} {\bibfnamefont {Jianshu}\
  \bibnamefont {Cao}},\ }\bibfield  {title} {\enquote {\bibinfo {title}
  {Coherent quantum transport in disordered systems: I. the influence of
  dephasing on the transport properties and absorption spectra on
  one-dimensional systems},}\ }\href {\doibase 10.1088/1367-2630/15/8/085010}
  {\bibfield  {journal} {\bibinfo  {journal} {New Journal of Physics}\ }\textbf
  {\bibinfo {volume} {15}},\ \bibinfo {pages} {085010} (\bibinfo {year}
  {2013})}\BibitemShut {NoStop}%
\bibitem [{\citenamefont {Wu}\ \emph {et~al.}(2013)\citenamefont {Wu},
  \citenamefont {Silbey},\ and\ \citenamefont {Cao}}]{CaoSilbeyWu2013}%
  \BibitemOpen
  \bibfield  {author} {\bibinfo {author} {\bibfnamefont {Jianlan}\ \bibnamefont
  {Wu}}, \bibinfo {author} {\bibfnamefont {Robert~J.}\ \bibnamefont {Silbey}},
  \ and\ \bibinfo {author} {\bibfnamefont {Jianshu}\ \bibnamefont {Cao}},\
  }\bibfield  {title} {\enquote {\bibinfo {title} {Generic mechanism of optimal
  energy transfer efficiency: A scaling theory of the mean first-passage time
  in exciton systems},}\ }\href {\doibase 10.1103/PhysRevLett.110.200402}
  {\bibfield  {journal} {\bibinfo  {journal} {Phys. Rev. Lett.}\ }\textbf
  {\bibinfo {volume} {110}},\ \bibinfo {pages} {200402} (\bibinfo {year}
  {2013})}\BibitemShut {NoStop}%
\bibitem [{\citenamefont {Zhang}\ \emph
  {et~al.}(2017{\natexlab{a}})\citenamefont {Zhang}, \citenamefont {Celardo},
  \citenamefont {Borgonovi},\ and\ \citenamefont
  {Kaplan}}]{Kaplan2017ExitSite}%
  \BibitemOpen
  \bibfield  {author} {\bibinfo {author} {\bibfnamefont {Yang}\ \bibnamefont
  {Zhang}}, \bibinfo {author} {\bibfnamefont {G.~Luca}\ \bibnamefont
  {Celardo}}, \bibinfo {author} {\bibfnamefont {Fausto}\ \bibnamefont
  {Borgonovi}}, \ and\ \bibinfo {author} {\bibfnamefont {Lev}\ \bibnamefont
  {Kaplan}},\ }\bibfield  {title} {\enquote {\bibinfo {title} {Opening-assisted
  coherent transport in the semiclassical regime},}\ }\href {\doibase
  10.1103/PhysRevE.95.022122} {\bibfield  {journal} {\bibinfo  {journal} {Phys.
  Rev. E}\ }\textbf {\bibinfo {volume} {95}},\ \bibinfo {pages} {022122}
  (\bibinfo {year} {2017}{\natexlab{a}})}\BibitemShut {NoStop}%
\bibitem [{\citenamefont {Zhang}\ \emph
  {et~al.}(2017{\natexlab{b}})\citenamefont {Zhang}, \citenamefont {Celardo},
  \citenamefont {Borgonovi},\ and\ \citenamefont {Kaplan}}]{Kaplan2017B}%
  \BibitemOpen
  \bibfield  {author} {\bibinfo {author} {\bibfnamefont {Yang}\ \bibnamefont
  {Zhang}}, \bibinfo {author} {\bibfnamefont {G.~Luca}\ \bibnamefont
  {Celardo}}, \bibinfo {author} {\bibfnamefont {Fausto}\ \bibnamefont
  {Borgonovi}}, \ and\ \bibinfo {author} {\bibfnamefont {Lev}\ \bibnamefont
  {Kaplan}},\ }\bibfield  {title} {\enquote {\bibinfo {title} {Optimal
  dephasing for ballistic energy transfer in disordered linear chains},}\
  }\href {\doibase 10.1103/PhysRevE.96.052103} {\bibfield  {journal} {\bibinfo
  {journal} {Phys. Rev. E}\ }\textbf {\bibinfo {volume} {96}},\ \bibinfo
  {pages} {052103} (\bibinfo {year} {2017}{\natexlab{b}})}\BibitemShut
  {NoStop}%
\bibitem [{\citenamefont {van Amerongen}\ \emph {et~al.}(2000)\citenamefont
  {van Amerongen}, \citenamefont {van Grondelle},\ and\ \citenamefont
  {Valkunas}}]{amerongen2000photosynthetic}%
  \BibitemOpen
  \bibfield  {author} {\bibinfo {author} {\bibfnamefont {Herbert}\ \bibnamefont
  {van Amerongen}}, \bibinfo {author} {\bibfnamefont {Rienk}\ \bibnamefont {van
  Grondelle}}, \ and\ \bibinfo {author} {\bibfnamefont {Leonas}\ \bibnamefont
  {Valkunas}},\ }\href {\doibase 10.1142/3609} {\emph {\bibinfo {title}
  {Photosynthetic Excitons}}}\ (\bibinfo  {publisher} {WORLD SCIENTIFIC},\
  \bibinfo {year} {2000})\BibitemShut {NoStop}%
\bibitem [{\citenamefont {Ritz}\ \emph {et~al.}(2002)\citenamefont {Ritz},
  \citenamefont {Damjanovi{\'c}},\ and\ \citenamefont
  {Schulten}}]{ritz2002quantum}%
  \BibitemOpen
  \bibfield  {author} {\bibinfo {author} {\bibfnamefont {Thorsten}\
  \bibnamefont {Ritz}}, \bibinfo {author} {\bibfnamefont {Ana}\ \bibnamefont
  {Damjanovi{\'c}}}, \ and\ \bibinfo {author} {\bibfnamefont {Klaus}\
  \bibnamefont {Schulten}},\ }\bibfield  {title} {\enquote {\bibinfo {title}
  {The quantum physics of photosynthesis},}\ }\href
  {https://onlinelibrary.wiley.com/doi/abs/10.1002/1439-7641%2820020315%293%3A3%3C243%3A%3AAID-CPHC243%3E3.0.CO%3B2-Y}
  {\bibfield  {journal} {\bibinfo  {journal} {ChemPhysChem}\ }\textbf {\bibinfo
  {volume} {3}},\ \bibinfo {pages} {243--248} (\bibinfo {year}
  {2002})}\BibitemShut {NoStop}%
\bibitem [{\citenamefont {Cheng}\ and\ \citenamefont
  {Fleming}(2009)}]{FlemingCheng2009}%
  \BibitemOpen
  \bibfield  {author} {\bibinfo {author} {\bibfnamefont {Yuan-Chung}\
  \bibnamefont {Cheng}}\ and\ \bibinfo {author} {\bibfnamefont {Graham~R.}\
  \bibnamefont {Fleming}},\ }\bibfield  {title} {\enquote {\bibinfo {title}
  {Dynamics of light harvesting in photosynthesis},}\ }\href {\doibase
  10.1146/annurev.physchem.040808.090259} {\bibfield  {journal} {\bibinfo
  {journal} {Annual Review of Physical Chemistry}\ }\textbf {\bibinfo {volume}
  {60}},\ \bibinfo {pages} {241--262} (\bibinfo {year} {2009})}\BibitemShut
  {NoStop}%
\bibitem [{\citenamefont {Plenio}\ and\ \citenamefont
  {Huelga}(2008)}]{plenio2008dephasing}%
  \BibitemOpen
  \bibfield  {author} {\bibinfo {author} {\bibfnamefont {Martin~B}\
  \bibnamefont {Plenio}}\ and\ \bibinfo {author} {\bibfnamefont {Susana~F}\
  \bibnamefont {Huelga}},\ }\bibfield  {title} {\enquote {\bibinfo {title}
  {Dephasing-assisted transport: quantum networks and biomolecules},}\ }\href
  {https://iopscience.iop.org/article/10.1088/1367-2630/10/11/113019/meta}
  {\bibfield  {journal} {\bibinfo  {journal} {New Journal of Physics}\ }\textbf
  {\bibinfo {volume} {10}},\ \bibinfo {pages} {113019} (\bibinfo {year}
  {2008})}\BibitemShut {NoStop}%
\bibitem [{\citenamefont {Rebentrost}\ \emph
  {et~al.}(2009{\natexlab{a}})\citenamefont {Rebentrost}, \citenamefont
  {Mohseni}, \citenamefont {Kassal}, \citenamefont {Lloyd},\ and\ \citenamefont
  {Aspuru-Guzik}}]{Rebentrost_2009}%
  \BibitemOpen
  \bibfield  {author} {\bibinfo {author} {\bibfnamefont {Patrick}\ \bibnamefont
  {Rebentrost}}, \bibinfo {author} {\bibfnamefont {Masoud}\ \bibnamefont
  {Mohseni}}, \bibinfo {author} {\bibfnamefont {Ivan}\ \bibnamefont {Kassal}},
  \bibinfo {author} {\bibfnamefont {Seth}\ \bibnamefont {Lloyd}}, \ and\
  \bibinfo {author} {\bibfnamefont {Al{\'{a}}n}\ \bibnamefont {Aspuru-Guzik}},\
  }\bibfield  {title} {\enquote {\bibinfo {title} {Environment-assisted quantum
  transport},}\ }\href {\doibase 10.1088/1367-2630/11/3/033003} {\bibfield
  {journal} {\bibinfo  {journal} {New Journal of Physics}\ }\textbf {\bibinfo
  {volume} {11}},\ \bibinfo {pages} {033003} (\bibinfo {year}
  {2009}{\natexlab{a}})}\BibitemShut {NoStop}%
\bibitem [{\citenamefont {Rebentrost}\ \emph
  {et~al.}(2009{\natexlab{b}})\citenamefont {Rebentrost}, \citenamefont
  {Mohseni},\ and\ \citenamefont {Aspuru-Guzik}}]{Alan2009}%
  \BibitemOpen
  \bibfield  {author} {\bibinfo {author} {\bibfnamefont {Patrick}\ \bibnamefont
  {Rebentrost}}, \bibinfo {author} {\bibfnamefont {Masoud}\ \bibnamefont
  {Mohseni}}, \ and\ \bibinfo {author} {\bibfnamefont {Alán}\ \bibnamefont
  {Aspuru-Guzik}},\ }\bibfield  {title} {\enquote {\bibinfo {title} {Role of
  quantum coherence and environmental fluctuations in chromophoric energy
  transport},}\ }\href {\doibase 10.1021/jp901724d} {\bibfield  {journal}
  {\bibinfo  {journal} {The Journal of Physical Chemistry B}\ }\textbf
  {\bibinfo {volume} {113}},\ \bibinfo {pages} {9942--9947} (\bibinfo {year}
  {2009}{\natexlab{b}})}\BibitemShut {NoStop}%
\bibitem [{\citenamefont {Sarovar}\ and\ \citenamefont
  {Whaley}(2013)}]{Sarovar_2013}%
  \BibitemOpen
  \bibfield  {author} {\bibinfo {author} {\bibfnamefont {Mohan}\ \bibnamefont
  {Sarovar}}\ and\ \bibinfo {author} {\bibfnamefont {K~Birgitta}\ \bibnamefont
  {Whaley}},\ }\bibfield  {title} {\enquote {\bibinfo {title} {Design
  principles and fundamental trade-offs in biomimetic light harvesting},}\
  }\href {\doibase 10.1088/1367-2630/15/1/013030} {\bibfield  {journal}
  {\bibinfo  {journal} {New Journal of Physics}\ }\textbf {\bibinfo {volume}
  {15}},\ \bibinfo {pages} {013030} (\bibinfo {year} {2013})}\BibitemShut
  {NoStop}%
\bibitem [{\citenamefont {Higgins}\ \emph {et~al.}(2014)\citenamefont
  {Higgins}, \citenamefont {Benjamin}, \citenamefont {Stace}, \citenamefont
  {Milburn}, \citenamefont {Lovett},\ and\ \citenamefont
  {Gauger}}]{higgins2014superabsorption}%
  \BibitemOpen
  \bibfield  {author} {\bibinfo {author} {\bibfnamefont {KDB}\ \bibnamefont
  {Higgins}}, \bibinfo {author} {\bibfnamefont {SC}~\bibnamefont {Benjamin}},
  \bibinfo {author} {\bibfnamefont {TM}~\bibnamefont {Stace}}, \bibinfo
  {author} {\bibfnamefont {GJ}~\bibnamefont {Milburn}}, \bibinfo {author}
  {\bibfnamefont {Brendon~William}\ \bibnamefont {Lovett}}, \ and\ \bibinfo
  {author} {\bibfnamefont {EM}~\bibnamefont {Gauger}},\ }\bibfield  {title}
  {\enquote {\bibinfo {title} {Superabsorption of light via quantum
  engineering},}\ }\href {https://www.nature.com/articles/ncomms5705}
  {\bibfield  {journal} {\bibinfo  {journal} {Nature communications}\ }\textbf
  {\bibinfo {volume} {5}},\ \bibinfo {pages} {4705} (\bibinfo {year}
  {2014})}\BibitemShut {NoStop}%
\bibitem [{\citenamefont {Celardo}\ \emph {et~al.}(2012)\citenamefont
  {Celardo}, \citenamefont {Borgonovi}, \citenamefont {Merkli}, \citenamefont
  {Tsifrinovich},\ and\ \citenamefont {Berman}}]{celardo2012superradiance}%
  \BibitemOpen
  \bibfield  {author} {\bibinfo {author} {\bibfnamefont {Giuseppe~L}\
  \bibnamefont {Celardo}}, \bibinfo {author} {\bibfnamefont {Fausto}\
  \bibnamefont {Borgonovi}}, \bibinfo {author} {\bibfnamefont {Marco}\
  \bibnamefont {Merkli}}, \bibinfo {author} {\bibfnamefont {Vladimir~I}\
  \bibnamefont {Tsifrinovich}}, \ and\ \bibinfo {author} {\bibfnamefont
  {Gennady~P}\ \bibnamefont {Berman}},\ }\bibfield  {title} {\enquote {\bibinfo
  {title} {Superradiance transition in photosynthetic light-harvesting
  complexes},}\ }\href {https://pubs.acs.org/doi/abs/10.1021/jp302627w}
  {\bibfield  {journal} {\bibinfo  {journal} {The Journal of Physical Chemistry
  C}\ }\textbf {\bibinfo {volume} {116}},\ \bibinfo {pages} {22105--22111}
  (\bibinfo {year} {2012})}\BibitemShut {NoStop}%
\bibitem [{\citenamefont {Park}\ \emph {et~al.}(2016)\citenamefont {Park},
  \citenamefont {Heldman}, \citenamefont {Rebentrost}, \citenamefont
  {Abbondanza}, \citenamefont {Iagatti}, \citenamefont {Alessi}, \citenamefont
  {Patrizi}, \citenamefont {Salvalaggio}, \citenamefont {Bussotti},
  \citenamefont {Mohseni} \emph {et~al.}}]{park2016enhanced}%
  \BibitemOpen
  \bibfield  {author} {\bibinfo {author} {\bibfnamefont {Heechul}\ \bibnamefont
  {Park}}, \bibinfo {author} {\bibfnamefont {Nimrod}\ \bibnamefont {Heldman}},
  \bibinfo {author} {\bibfnamefont {Patrick}\ \bibnamefont {Rebentrost}},
  \bibinfo {author} {\bibfnamefont {Luigi}\ \bibnamefont {Abbondanza}},
  \bibinfo {author} {\bibfnamefont {Alessandro}\ \bibnamefont {Iagatti}},
  \bibinfo {author} {\bibfnamefont {Andrea}\ \bibnamefont {Alessi}}, \bibinfo
  {author} {\bibfnamefont {Barbara}\ \bibnamefont {Patrizi}}, \bibinfo {author}
  {\bibfnamefont {Mario}\ \bibnamefont {Salvalaggio}}, \bibinfo {author}
  {\bibfnamefont {Laura}\ \bibnamefont {Bussotti}}, \bibinfo {author}
  {\bibfnamefont {Masoud}\ \bibnamefont {Mohseni}},  \emph {et~al.},\
  }\bibfield  {title} {\enquote {\bibinfo {title} {Enhanced energy transport in
  genetically engineered excitonic networks},}\ }\href
  {https://www.nature.com/articles/nmat4448} {\bibfield  {journal} {\bibinfo
  {journal} {Nature materials}\ }\textbf {\bibinfo {volume} {15}},\ \bibinfo
  {pages} {211} (\bibinfo {year} {2016})}\BibitemShut {NoStop}%
\bibitem [{\citenamefont {Caldeira}\ and\ \citenamefont
  {Leggett}(1983{\natexlab{a}})}]{Caldeira1983}%
  \BibitemOpen
  \bibfield  {author} {\bibinfo {author} {\bibfnamefont {Amir~O}\ \bibnamefont
  {Caldeira}}\ and\ \bibinfo {author} {\bibfnamefont {Anthony~J}\ \bibnamefont
  {Leggett}},\ }\bibfield  {title} {\enquote {\bibinfo {title} {Path integral
  approach to quantum brownian motion},}\ }\href
  {http://www.sciencedirect.com/science/article/pii/0378437183900134}
  {\bibfield  {journal} {\bibinfo  {journal} {Physica A: Statistical mechanics
  and its Applications}\ }\textbf {\bibinfo {volume} {121}},\ \bibinfo {pages}
  {587--616} (\bibinfo {year} {1983}{\natexlab{a}})}\BibitemShut {NoStop}%
\bibitem [{\citenamefont {Caldeira}\ and\ \citenamefont
  {Leggett}(1983{\natexlab{b}})}]{CALDEIRA1983374}%
  \BibitemOpen
  \bibfield  {author} {\bibinfo {author} {\bibfnamefont {Amir~O}\ \bibnamefont
  {Caldeira}}\ and\ \bibinfo {author} {\bibfnamefont {Anthony~J}\ \bibnamefont
  {Leggett}},\ }\bibfield  {title} {\enquote {\bibinfo {title} {Quantum
  tunnelling in a dissipative system},}\ }\href {\doibase
  https://doi.org/10.1016/0003-4916(83)90202-6} {\bibfield  {journal} {\bibinfo
   {journal} {Annals of Physics}\ }\textbf {\bibinfo {volume} {149}},\ \bibinfo
  {pages} {374 -- 456} (\bibinfo {year} {1983}{\natexlab{b}})}\BibitemShut
  {NoStop}%
\bibitem [{\citenamefont {Cohen}(1997)}]{Cohen1997}%
  \BibitemOpen
  \bibfield  {author} {\bibinfo {author} {\bibfnamefont {Doron}\ \bibnamefont
  {Cohen}},\ }\bibfield  {title} {\enquote {\bibinfo {title} {Unified model for
  the study of diffusion localization and dissipation},}\ }\href
  {https://journals.aps.org/pre/abstract/10.1103/PhysRevE.55.1422} {\bibfield
  {journal} {\bibinfo  {journal} {Physical Review E}\ }\textbf {\bibinfo
  {volume} {55}},\ \bibinfo {pages} {1422} (\bibinfo {year}
  {1997})}\BibitemShut {NoStop}%
\bibitem [{\citenamefont {Esposito}\ and\ \citenamefont
  {Gaspard}(2005)}]{EspositoGaspard2005}%
  \BibitemOpen
  \bibfield  {author} {\bibinfo {author} {\bibfnamefont {Massimiliano}\
  \bibnamefont {Esposito}}\ and\ \bibinfo {author} {\bibfnamefont {Pierre}\
  \bibnamefont {Gaspard}},\ }\bibfield  {title} {\enquote {\bibinfo {title}
  {Emergence of diffusion in finite quantum systems},}\ }\href
  {https://link.springer.com/article/10.1007/s10955-005-7577-x} {\bibfield
  {journal} {\bibinfo  {journal} {Journal of statistical physics}\ }\textbf
  {\bibinfo {volume} {121}},\ \bibinfo {pages} {463--496} (\bibinfo {year}
  {2005})}\BibitemShut {NoStop}%
\bibitem [{\citenamefont {Dyson}(1953)}]{Dyson1953}%
  \BibitemOpen
  \bibfield  {author} {\bibinfo {author} {\bibfnamefont {Freeman~J.}\
  \bibnamefont {Dyson}},\ }\bibfield  {title} {\enquote {\bibinfo {title} {The
  dynamics of a disordered linear chain},}\ }\href
  {https://link.aps.org/doi/10.1103/PhysRev.92.1331} {\bibfield  {journal}
  {\bibinfo  {journal} {Phys. Rev.}\ }\textbf {\bibinfo {volume} {92}},\
  \bibinfo {pages} {1331--1338} (\bibinfo {year} {1953})}\BibitemShut {NoStop}%
\bibitem [{\citenamefont {Sinai}(1983)}]{Sinai1983}%
  \BibitemOpen
  \bibfield  {author} {\bibinfo {author} {\bibfnamefont {Ya.~G.}\ \bibnamefont
  {Sinai}},\ }\bibfield  {title} {\enquote {\bibinfo {title} {The limiting
  behavior of a one-dimensional random walk in a random medium},}\ }\href
  {https://doi.org/10.1137/1127028} {\bibfield  {journal} {\bibinfo  {journal}
  {Theory of Probability \& Its Applications}\ }\textbf {\bibinfo {volume}
  {27}},\ \bibinfo {pages} {256--268} (\bibinfo {year} {1983})}\BibitemShut
  {NoStop}%
\bibitem [{\citenamefont {Derrida}\ and\ \citenamefont
  {Pomeau}(1982)}]{DerridaPomeau1982}%
  \BibitemOpen
  \bibfield  {author} {\bibinfo {author} {\bibfnamefont {B}~\bibnamefont
  {Derrida}}\ and\ \bibinfo {author} {\bibfnamefont {Ya}~\bibnamefont
  {Pomeau}},\ }\bibfield  {title} {\enquote {\bibinfo {title} {Classical
  diffusion on a random chain},}\ }\href
  {http://journals.aps.org/prl/abstract/10.1103/PhysRevLett.48.627} {\bibfield
  {journal} {\bibinfo  {journal} {Physical Review Letters}\ }\textbf {\bibinfo
  {volume} {48}},\ \bibinfo {pages} {627} (\bibinfo {year} {1982})}\BibitemShut
  {NoStop}%
\bibitem [{\citenamefont {Derrida}(1983)}]{Derrida1983}%
  \BibitemOpen
  \bibfield  {author} {\bibinfo {author} {\bibfnamefont {Bernard}\ \bibnamefont
  {Derrida}},\ }\bibfield  {title} {\enquote {\bibinfo {title} {Velocity and
  diffusion constant of a periodic one-dimensional hopping model},}\ }\href
  {https://doi.org/10.1007/BF01019492} {\bibfield  {journal} {\bibinfo
  {journal} {Journal of Statistical Physics}\ }\textbf {\bibinfo {volume}
  {31}},\ \bibinfo {pages} {433--450} (\bibinfo {year} {1983})}\BibitemShut
  {NoStop}%
\bibitem [{\citenamefont {Havlin}\ and\ \citenamefont
  {Ben-Avraham}(1987)}]{havlin1987diffusion}%
  \BibitemOpen
  \bibfield  {author} {\bibinfo {author} {\bibfnamefont {Shlomo}\ \bibnamefont
  {Havlin}}\ and\ \bibinfo {author} {\bibfnamefont {Daniel}\ \bibnamefont
  {Ben-Avraham}},\ }\bibfield  {title} {\enquote {\bibinfo {title} {Diffusion
  in disordered media},}\ }\href
  {https://www.tandfonline.com/doi/abs/10.1080/00018738700101072} {\bibfield
  {journal} {\bibinfo  {journal} {Advances in Physics}\ }\textbf {\bibinfo
  {volume} {36}},\ \bibinfo {pages} {695--798} (\bibinfo {year}
  {1987})}\BibitemShut {NoStop}%
\bibitem [{\citenamefont {Bouchaud}\ \emph {et~al.}(1990)\citenamefont
  {Bouchaud}, \citenamefont {Comtet}, \citenamefont {Georges},\ and\
  \citenamefont {Le~Doussal}}]{Bouchaud1990}%
  \BibitemOpen
  \bibfield  {author} {\bibinfo {author} {\bibfnamefont {J-Ph}\ \bibnamefont
  {Bouchaud}}, \bibinfo {author} {\bibfnamefont {A}~\bibnamefont {Comtet}},
  \bibinfo {author} {\bibfnamefont {A}~\bibnamefont {Georges}}, \ and\ \bibinfo
  {author} {\bibfnamefont {P}~\bibnamefont {Le~Doussal}},\ }\bibfield  {title}
  {\enquote {\bibinfo {title} {Classical diffusion of a particle in a
  one-dimensional random force field},}\ }\href
  {http://www.sciencedirect.com/science/article/pii/000349169090043N}
  {\bibfield  {journal} {\bibinfo  {journal} {Annals of Physics}\ }\textbf
  {\bibinfo {volume} {201}},\ \bibinfo {pages} {285--341} (\bibinfo {year}
  {1990})}\BibitemShut {NoStop}%
\bibitem [{\citenamefont {Bouchaud}\ and\ \citenamefont
  {Georges}(1990)}]{BOUCHAUD1990a}%
  \BibitemOpen
  \bibfield  {author} {\bibinfo {author} {\bibfnamefont {Jean-Philippe}\
  \bibnamefont {Bouchaud}}\ and\ \bibinfo {author} {\bibfnamefont {Antoine}\
  \bibnamefont {Georges}},\ }\bibfield  {title} {\enquote {\bibinfo {title}
  {Anomalous diffusion in disordered media: Statistical mechanisms, models and
  physical applications},}\ }\href
  {http://www.sciencedirect.com/science/article/pii/037015739090099N}
  {\bibfield  {journal} {\bibinfo  {journal} {Physics Reports}\ }\textbf
  {\bibinfo {volume} {195}},\ \bibinfo {pages} {127 -- 293} (\bibinfo {year}
  {1990})}\BibitemShut {NoStop}%
\bibitem [{\citenamefont {Hurowitz}\ and\ \citenamefont
  {Cohen}(2016{\natexlab{a}})}]{HurowitzCohen2016}%
  \BibitemOpen
  \bibfield  {author} {\bibinfo {author} {\bibfnamefont {Daniel}\ \bibnamefont
  {Hurowitz}}\ and\ \bibinfo {author} {\bibfnamefont {Doron}\ \bibnamefont
  {Cohen}},\ }\bibfield  {title} {\enquote {\bibinfo {title} {Percolation,
  sliding, localization and relaxation in topologically closed circuits},}\
  }\href {http://www.nature.com/articles/srep22735} {\bibfield  {journal}
  {\bibinfo  {journal} {Scientific reports}\ }\textbf {\bibinfo {volume} {6}}
  (\bibinfo {year} {2016}{\natexlab{a}})}\BibitemShut {NoStop}%
\bibitem [{\citenamefont {Hurowitz}\ and\ \citenamefont
  {Cohen}(2016{\natexlab{b}})}]{HurowitzCohen2016a}%
  \BibitemOpen
  \bibfield  {author} {\bibinfo {author} {\bibfnamefont {Daniel}\ \bibnamefont
  {Hurowitz}}\ and\ \bibinfo {author} {\bibfnamefont {Doron}\ \bibnamefont
  {Cohen}},\ }\bibfield  {title} {\enquote {\bibinfo {title} {Relaxation rate
  of a stochastic spreading process in a closed ring},}\ }\href
  {https://link.aps.org/doi/10.1103/PhysRevE.93.062143} {\bibfield  {journal}
  {\bibinfo  {journal} {Phys. Rev. E}\ }\textbf {\bibinfo {volume} {93}},\
  \bibinfo {pages} {062143} (\bibinfo {year} {2016}{\natexlab{b}})}\BibitemShut
  {NoStop}%
\bibitem [{\citenamefont {Hatano}\ and\ \citenamefont
  {Nelson}(1996)}]{Hatano1996}%
  \BibitemOpen
  \bibfield  {author} {\bibinfo {author} {\bibfnamefont {Naomichi}\
  \bibnamefont {Hatano}}\ and\ \bibinfo {author} {\bibfnamefont {David~R.}\
  \bibnamefont {Nelson}},\ }\bibfield  {title} {\enquote {\bibinfo {title}
  {Localization transitions in non-hermitian quantum mechanics},}\ }\href
  {https://link.aps.org/doi/10.1103/PhysRevLett.77.570} {\bibfield  {journal}
  {\bibinfo  {journal} {Phys. Rev. Lett.}\ }\textbf {\bibinfo {volume} {77}},\
  \bibinfo {pages} {570--573} (\bibinfo {year} {1996})}\BibitemShut {NoStop}%
\bibitem [{\citenamefont {Hatano}\ and\ \citenamefont
  {Nelson}(1997)}]{Hatano1997}%
  \BibitemOpen
  \bibfield  {author} {\bibinfo {author} {\bibfnamefont {Naomichi}\
  \bibnamefont {Hatano}}\ and\ \bibinfo {author} {\bibfnamefont {David~R.}\
  \bibnamefont {Nelson}},\ }\bibfield  {title} {\enquote {\bibinfo {title}
  {Vortex pinning and non-hermitian quantum mechanics},}\ }\href
  {https://link.aps.org/doi/10.1103/PhysRevB.56.8651} {\bibfield  {journal}
  {\bibinfo  {journal} {Phys. Rev. B}\ }\textbf {\bibinfo {volume} {56}},\
  \bibinfo {pages} {8651--8673} (\bibinfo {year} {1997})}\BibitemShut {NoStop}%
\bibitem [{\citenamefont {Hatano}(1998)}]{Hatano1998}%
  \BibitemOpen
  \bibfield  {author} {\bibinfo {author} {\bibfnamefont {Naomichi}\
  \bibnamefont {Hatano}},\ }\bibfield  {title} {\enquote {\bibinfo {title}
  {Localization in non-hermitian quantum mechanics and flux-line pinning in
  superconductors},}\ }\href@noop {} {\bibfield  {journal} {\bibinfo  {journal}
  {Physica A: Statistical Mechanics and its Applications}\ }\textbf {\bibinfo
  {volume} {254}},\ \bibinfo {pages} {317--331} (\bibinfo {year}
  {1998})}\BibitemShut {NoStop}%
\bibitem [{\citenamefont {K.~Lubensky}\ and\ \citenamefont
  {R.~Nelson}(2000)}]{LubenskyNelson2000}%
  \BibitemOpen
  \bibfield  {author} {\bibinfo {author} {\bibfnamefont {David}\ \bibnamefont
  {K.~Lubensky}}\ and\ \bibinfo {author} {\bibfnamefont {David}\ \bibnamefont
  {R.~Nelson}},\ }\bibfield  {title} {\enquote {\bibinfo {title} {Pulling
  pinned polymers and unzipping dna},}\ }\href {\doibase
  10.1103/PhysRevLett.85.1572} {\bibfield  {journal} {\bibinfo  {journal}
  {Physical review letters}\ }\textbf {\bibinfo {volume} {85}},\ \bibinfo
  {pages} {1572--5} (\bibinfo {year} {2000})}\BibitemShut {NoStop}%
\bibitem [{\citenamefont {Lubensky}\ and\ \citenamefont
  {Nelson}(2002)}]{LubenskyNelson2002}%
  \BibitemOpen
  \bibfield  {author} {\bibinfo {author} {\bibfnamefont {David~K.}\
  \bibnamefont {Lubensky}}\ and\ \bibinfo {author} {\bibfnamefont {David~R.}\
  \bibnamefont {Nelson}},\ }\bibfield  {title} {\enquote {\bibinfo {title}
  {Single molecule statistics and the polynucleotide unzipping transition},}\
  }\href {\doibase 10.1103/PhysRevE.65.031917} {\bibfield  {journal} {\bibinfo
  {journal} {Phys. Rev. E}\ }\textbf {\bibinfo {volume} {65}},\ \bibinfo
  {pages} {031917} (\bibinfo {year} {2002})}\BibitemShut {NoStop}%
\bibitem [{\citenamefont {Amir}\ \emph {et~al.}(2015)\citenamefont {Amir},
  \citenamefont {Hatano},\ and\ \citenamefont {Nelson}}]{Amir2015}%
  \BibitemOpen
  \bibfield  {author} {\bibinfo {author} {\bibfnamefont {Ariel}\ \bibnamefont
  {Amir}}, \bibinfo {author} {\bibfnamefont {Naomichi}\ \bibnamefont {Hatano}},
  \ and\ \bibinfo {author} {\bibfnamefont {David~R}\ \bibnamefont {Nelson}},\
  }\bibfield  {title} {\enquote {\bibinfo {title} {Localization in
  non-hermitian chains with excitatory/inhibitory connections},}\ }\href@noop
  {} {\bibfield  {journal} {\bibinfo  {journal} {arXiv preprint
  arXiv:1512.05478}\ } (\bibinfo {year} {2015})}\BibitemShut {NoStop}%
\bibitem [{\citenamefont {Amir}\ \emph {et~al.}(2016)\citenamefont {Amir},
  \citenamefont {Hatano},\ and\ \citenamefont {Nelson}}]{Amir2016}%
  \BibitemOpen
  \bibfield  {author} {\bibinfo {author} {\bibfnamefont {Ariel}\ \bibnamefont
  {Amir}}, \bibinfo {author} {\bibfnamefont {Naomichi}\ \bibnamefont {Hatano}},
  \ and\ \bibinfo {author} {\bibfnamefont {David~R}\ \bibnamefont {Nelson}},\
  }\bibfield  {title} {\enquote {\bibinfo {title} {Non-hermitian localization
  in biological networks},}\ }\href
  {https://journals.aps.org/pre/abstract/10.1103/PhysRevE.93.042310} {\bibfield
   {journal} {\bibinfo  {journal} {Physical Review E}\ }\textbf {\bibinfo
  {volume} {93}},\ \bibinfo {pages} {042310} (\bibinfo {year}
  {2016})}\BibitemShut {NoStop}%
\bibitem [{\citenamefont {Dahmen}\ \emph {et~al.}(1999)\citenamefont {Dahmen},
  \citenamefont {Nelson},\ and\ \citenamefont {Shnerb}}]{Shnerb99}%
  \BibitemOpen
  \bibfield  {author} {\bibinfo {author} {\bibfnamefont {Karin~A.}\
  \bibnamefont {Dahmen}}, \bibinfo {author} {\bibfnamefont {David~R.}\
  \bibnamefont {Nelson}}, \ and\ \bibinfo {author} {\bibfnamefont {Nadav~M.}\
  \bibnamefont {Shnerb}},\ }\bibfield  {title} {\enquote {\bibinfo {title}
  {{P}opulation dynamics and non-{H}ermitian localization},}\ }in\ \href
  {https://link.springer.com/chapter/10.1007/BFb0105012} {\emph {\bibinfo
  {booktitle} {Statistical Mechanics of Biocomplexity}}},\ \bibinfo {editor}
  {edited by\ \bibinfo {editor} {\bibfnamefont {D.}~\bibnamefont {Reguera}},
  \bibinfo {editor} {\bibfnamefont {J.M.G.}\ \bibnamefont {Vilar}}, \ and\
  \bibinfo {editor} {\bibfnamefont {J.M.}\ \bibnamefont {Rub{\'i}}}}\ (\bibinfo
   {publisher} {Springer Berlin Heidelberg},\ \bibinfo {address} {Berlin,
  Heidelberg},\ \bibinfo {year} {1999})\ pp.\ \bibinfo {pages}
  {124--151}\BibitemShut {NoStop}%
\bibitem [{\citenamefont {Rivas}\ and\ \citenamefont
  {Huelga}(2012)}]{Rivas2012}%
  \BibitemOpen
  \bibfield  {author} {\bibinfo {author} {\bibfnamefont {{\'A}ngel}\
  \bibnamefont {Rivas}}\ and\ \bibinfo {author} {\bibfnamefont {Susana~F}\
  \bibnamefont {Huelga}},\ }\href@noop {} {\emph {\bibinfo {title} {Open
  Quantum Systems}}}\ (\bibinfo  {publisher} {Springer},\ \bibinfo {year}
  {2012})\BibitemShut {NoStop}%
\bibitem [{\citenamefont {Shapira}\ and\ \citenamefont {Cohen}()}]{qssXs-prep}%
  \BibitemOpen
  \bibfield  {author} {\bibinfo {author} {\bibfnamefont {Dekel}\ \bibnamefont
  {Shapira}}\ and\ \bibinfo {author} {\bibfnamefont {Doron}\ \bibnamefont
  {Cohen}},\ }\href@noop {} {\bibinfo  {journal} {In preparation}\
  }\BibitemShut {NoStop}%
\bibitem [{\citenamefont {Hakim}\ and\ \citenamefont
  {Ambegaokar}(1985)}]{HakimAmbegaokar1985}%
  \BibitemOpen
\bibfield  {journal} {  }\bibfield  {author} {\bibinfo {author} {\bibfnamefont
  {Vincent}\ \bibnamefont {Hakim}}\ and\ \bibinfo {author} {\bibfnamefont
  {Vinay}\ \bibnamefont {Ambegaokar}},\ }\bibfield  {title} {\enquote {\bibinfo
  {title} {Quantum theory of a free particle interacting with a linearly
  dissipative environment},}\ }\href {\doibase 10.1103/PhysRevA.32.423}
  {\bibfield  {journal} {\bibinfo  {journal} {Phys. Rev. A}\ }\textbf {\bibinfo
  {volume} {32}},\ \bibinfo {pages} {423--434} (\bibinfo {year}
  {1985})}\BibitemShut {NoStop}%
\bibitem [{\citenamefont {Leggett}\ \emph {et~al.}(1987)\citenamefont
  {Leggett}, \citenamefont {Chakravarty}, \citenamefont {Dorsey}, \citenamefont
  {Fisher}, \citenamefont {Garg},\ and\ \citenamefont
  {Zwerger}}]{LeggettEtAlDynamicsTwoLevel1987}%
  \BibitemOpen
  \bibfield  {author} {\bibinfo {author} {\bibfnamefont {A.~J.}\ \bibnamefont
  {Leggett}}, \bibinfo {author} {\bibfnamefont {S.}~\bibnamefont
  {Chakravarty}}, \bibinfo {author} {\bibfnamefont {A.~T.}\ \bibnamefont
  {Dorsey}}, \bibinfo {author} {\bibfnamefont {Matthew P.~A.}\ \bibnamefont
  {Fisher}}, \bibinfo {author} {\bibfnamefont {Anupam}\ \bibnamefont {Garg}}, \
  and\ \bibinfo {author} {\bibfnamefont {W.}~\bibnamefont {Zwerger}},\
  }\bibfield  {title} {\enquote {\bibinfo {title} {Dynamics of the dissipative
  two-state system},}\ }\href {\doibase 10.1103/RevModPhys.59.1} {\bibfield
  {journal} {\bibinfo  {journal} {Rev. Mod. Phys.}\ }\textbf {\bibinfo {volume}
  {59}},\ \bibinfo {pages} {1--85} (\bibinfo {year} {1987})}\BibitemShut
  {NoStop}%
\bibitem [{\citenamefont {Aslangul}\ \emph {et~al.}(1986)\citenamefont
  {Aslangul}, \citenamefont {Pottier},\ and\ \citenamefont
  {Saint-James}}]{aslangul1986quantum}%
  \BibitemOpen
  \bibfield  {author} {\bibinfo {author} {\bibfnamefont {C}~\bibnamefont
  {Aslangul}}, \bibinfo {author} {\bibfnamefont {N}~\bibnamefont {Pottier}}, \
  and\ \bibinfo {author} {\bibfnamefont {D}~\bibnamefont {Saint-James}},\
  }\bibfield  {title} {\enquote {\bibinfo {title} {Quantum ohmic dissipation:
  cross-over between quantum tunnelling and thermally resisted motion in a
  biased tight-binding lattice},}\ }\href
  {https://hal.archives-ouvertes.fr/jpa-00210364/document} {\bibfield
  {journal} {\bibinfo  {journal} {Journal de Physique}\ }\textbf {\bibinfo
  {volume} {47}},\ \bibinfo {pages} {1671--1685} (\bibinfo {year}
  {1986})}\BibitemShut {NoStop}%
\bibitem [{\citenamefont {{Aslangul, C.}}\ \emph {et~al.}(1987)\citenamefont
  {{Aslangul, C.}}, \citenamefont {{Pottier, N.}},\ and\ \citenamefont
  {{Saint-James, D.}}}]{AslangulPeriodicPotential1987}%
  \BibitemOpen
  \bibfield  {author} {\bibinfo {author} {\bibnamefont {{Aslangul, C.}}},
  \bibinfo {author} {\bibnamefont {{Pottier, N.}}}, \ and\ \bibinfo {author}
  {\bibnamefont {{Saint-James, D.}}},\ }\bibfield  {title} {\enquote {\bibinfo
  {title} {Quantum brownian motion in a periodic potential: a pedestrian
  approach},}\ }\href {\doibase 10.1051/jphys:019870048070109300} {\bibfield
  {journal} {\bibinfo  {journal} {J. Phys. France}\ }\textbf {\bibinfo {volume}
  {48}},\ \bibinfo {pages} {1093--1110} (\bibinfo {year} {1987})}\BibitemShut
  {NoStop}%
\bibitem [{\citenamefont {Fisher}\ and\ \citenamefont
  {Zwerger}(1985)}]{Fisher1985QuantumBrownianPeriodic}%
  \BibitemOpen
  \bibfield  {author} {\bibinfo {author} {\bibfnamefont {Matthew P.~A.}\
  \bibnamefont {Fisher}}\ and\ \bibinfo {author} {\bibfnamefont {Wilhelm}\
  \bibnamefont {Zwerger}},\ }\bibfield  {title} {\enquote {\bibinfo {title}
  {Quantum brownian motion in a periodic potential},}\ }\href {\doibase
  10.1103/PhysRevB.32.6190} {\bibfield  {journal} {\bibinfo  {journal} {Phys.
  Rev. B}\ }\textbf {\bibinfo {volume} {32}},\ \bibinfo {pages} {6190--6206}
  (\bibinfo {year} {1985})}\BibitemShut {NoStop}%
\bibitem [{\citenamefont {Schwinger}(1961)}]{Schwinger1961}%
  \BibitemOpen
  \bibfield  {author} {\bibinfo {author} {\bibfnamefont {Julian}\ \bibnamefont
  {Schwinger}},\ }\bibfield  {title} {\enquote {\bibinfo {title} {Brownian
  motion of a quantum oscillator},}\ }\href {\doibase 10.1063/1.1703727}
  {\bibfield  {journal} {\bibinfo  {journal} {Journal of Mathematical Physics}\
  }\textbf {\bibinfo {volume} {2}},\ \bibinfo {pages} {407--432} (\bibinfo
  {year} {1961})}\BibitemShut {NoStop}%
\bibitem [{\citenamefont {Grabert}\ \emph {et~al.}(1988)\citenamefont
  {Grabert}, \citenamefont {Schramm},\ and\ \citenamefont
  {Ingold}}]{GRABERT1988115}%
  \BibitemOpen
  \bibfield  {author} {\bibinfo {author} {\bibfnamefont {Hermann}\ \bibnamefont
  {Grabert}}, \bibinfo {author} {\bibfnamefont {Peter}\ \bibnamefont
  {Schramm}}, \ and\ \bibinfo {author} {\bibfnamefont {Gert-Ludwig}\
  \bibnamefont {Ingold}},\ }\bibfield  {title} {\enquote {\bibinfo {title}
  {Quantum brownian motion: The functional integral approach},}\ }\href
  {\doibase https://doi.org/10.1016/0370-1573(88)90023-3} {\bibfield  {journal}
  {\bibinfo  {journal} {Physics Reports}\ }\textbf {\bibinfo {volume} {168}},\
  \bibinfo {pages} {115 -- 207} (\bibinfo {year} {1988})}\BibitemShut {NoStop}%
\bibitem [{\citenamefont {Hänggi}\ and\ \citenamefont
  {Ingold}(2005)}]{HanggiQBM2005}%
  \BibitemOpen
  \bibfield  {author} {\bibinfo {author} {\bibfnamefont {Peter}\ \bibnamefont
  {Hänggi}}\ and\ \bibinfo {author} {\bibfnamefont {Gert-Ludwig}\ \bibnamefont
  {Ingold}},\ }\bibfield  {title} {\enquote {\bibinfo {title} {Fundamental
  aspects of quantum brownian motion},}\ }\href {\doibase 10.1063/1.1853631}
  {\bibfield  {journal} {\bibinfo  {journal} {Chaos: An Interdisciplinary
  Journal of Nonlinear Science}\ }\textbf {\bibinfo {volume} {15}},\ \bibinfo
  {pages} {026105} (\bibinfo {year} {2005})}\BibitemShut {NoStop}%
\bibitem [{\citenamefont {Schmid}(1983)}]{Schmid1983}%
  \BibitemOpen
  \bibfield  {author} {\bibinfo {author} {\bibfnamefont {Albert}\ \bibnamefont
  {Schmid}},\ }\bibfield  {title} {\enquote {\bibinfo {title} {Diffusion and
  localization in a dissipative quantum system},}\ }\href
  {http://journals.aps.org/prl/pdf/10.1103/PhysRevLett.51.1506} {\bibfield
  {journal} {\bibinfo  {journal} {Physical Review Letters}\ }\textbf {\bibinfo
  {volume} {51}},\ \bibinfo {pages} {1506} (\bibinfo {year}
  {1983})}\BibitemShut {NoStop}%
\bibitem [{\citenamefont {Weiss}\ \emph {et~al.}(1991)\citenamefont {Weiss},
  \citenamefont {Sassetti}, \citenamefont {Negele},\ and\ \citenamefont
  {Wollensak}}]{Weiss1991}%
  \BibitemOpen
  \bibfield  {author} {\bibinfo {author} {\bibfnamefont {Ulrich}\ \bibnamefont
  {Weiss}}, \bibinfo {author} {\bibfnamefont {Maura}\ \bibnamefont {Sassetti}},
  \bibinfo {author} {\bibfnamefont {Thomas}\ \bibnamefont {Negele}}, \ and\
  \bibinfo {author} {\bibfnamefont {Matthias}\ \bibnamefont {Wollensak}},\
  }\bibfield  {title} {\enquote {\bibinfo {title} {Dissipative quantum dynamics
  in a multiwell system},}\ }\href {https://doi.org/10.1007/BF01314023}
  {\bibfield  {journal} {\bibinfo  {journal} {Zeitschrift f{\"u}r Physik B
  Condensed Matter}\ }\textbf {\bibinfo {volume} {84}},\ \bibinfo {pages}
  {471--482} (\bibinfo {year} {1991})}\BibitemShut {NoStop}%
\bibitem [{\citenamefont
  {{\v{Z}}nidari{\v{c}}}(2010)}]{znidaricOpenQuantumChain2010}%
  \BibitemOpen
  \bibfield  {author} {\bibinfo {author} {\bibfnamefont {Marko}\ \bibnamefont
  {{\v{Z}}nidari{\v{c}}}},\ }\bibfield  {title} {\enquote {\bibinfo {title}
  {Exact solution for a diffusive nonequilibrium steady state of an open
  quantum chain},}\ }\href {\doibase 10.1088/1742-5468/2010/05/l05002}
  {\bibfield  {journal} {\bibinfo  {journal} {Journal of Statistical Mechanics:
  Theory and Experiment}\ }\textbf {\bibinfo {volume} {2010}},\ \bibinfo
  {pages} {L05002} (\bibinfo {year} {2010})}\BibitemShut {NoStop}%
\bibitem [{\citenamefont
  {Eisler}(2011)}]{EislerCrossoverBallistiDiffusiveQuantum_2011}%
  \BibitemOpen
  \bibfield  {author} {\bibinfo {author} {\bibfnamefont {Viktor}\ \bibnamefont
  {Eisler}},\ }\bibfield  {title} {\enquote {\bibinfo {title} {Crossover
  between ballistic and diffusive transport: the quantum exclusion process},}\
  }\href {\doibase 10.1088/1742-5468/2011/06/p06007} {\bibfield  {journal}
  {\bibinfo  {journal} {Journal of Statistical Mechanics: Theory and
  Experiment}\ }\textbf {\bibinfo {volume} {2011}},\ \bibinfo {pages} {P06007}
  (\bibinfo {year} {2011})}\BibitemShut {NoStop}%
\bibitem [{\citenamefont {{\v{Z}}nidari{\v{c}}}\ and\ \citenamefont
  {Horvat}(2013)}]{znidaricDisorderDephasing2013}%
  \BibitemOpen
  \bibfield  {author} {\bibinfo {author} {\bibfnamefont {Marko}\ \bibnamefont
  {{\v{Z}}nidari{\v{c}}}}\ and\ \bibinfo {author} {\bibfnamefont {Martin}\
  \bibnamefont {Horvat}},\ }\bibfield  {title} {\enquote {\bibinfo {title}
  {Transport in a disordered tight-binding chain with dephasing},}\ }\href
  {\doibase 10.1140/epjb/e2012-30730-9} {\bibfield  {journal} {\bibinfo
  {journal} {The European Physical Journal B}\ }\textbf {\bibinfo {volume}
  {86}},\ \bibinfo {pages} {67} (\bibinfo {year} {2013})}\BibitemShut {NoStop}%
\bibitem [{\citenamefont {Han}\ and\ \citenamefont
  {Hartnoll}(2018)}]{Han_local_bath_diffusion_2018}%
  \BibitemOpen
  \bibfield  {author} {\bibinfo {author} {\bibfnamefont {Xizhi}\ \bibnamefont
  {Han}}\ and\ \bibinfo {author} {\bibfnamefont {Sean~A.}\ \bibnamefont
  {Hartnoll}},\ }\bibfield  {title} {\enquote {\bibinfo {title} {Locality bound
  for dissipative quantum transport},}\ }\href {\doibase
  10.1103/PhysRevLett.121.170601} {\bibfield  {journal} {\bibinfo  {journal}
  {Phys. Rev. Lett.}\ }\textbf {\bibinfo {volume} {121}},\ \bibinfo {pages}
  {170601} (\bibinfo {year} {2018})}\BibitemShut {NoStop}%
\bibitem [{\citenamefont {Breuer}\ and\ \citenamefont
  {Petruccione}(2002)}]{Breuer2002}%
  \BibitemOpen
  \bibfield  {author} {\bibinfo {author} {\bibfnamefont {Heinz-Peter}\
  \bibnamefont {Breuer}}\ and\ \bibinfo {author} {\bibfnamefont {Francesco}\
  \bibnamefont {Petruccione}},\ }\href@noop {} {\emph {\bibinfo {title} {The
  theory of open quantum systems}}}\ (\bibinfo  {publisher} {Oxford University
  Press on Demand},\ \bibinfo {year} {2002})\BibitemShut {NoStop}%
\bibitem [{\citenamefont {Davies}(1976)}]{davies1976quantum}%
  \BibitemOpen
  \bibfield  {author} {\bibinfo {author} {\bibfnamefont {Edward~Brian}\
  \bibnamefont {Davies}},\ }\href@noop {} {\emph {\bibinfo {title} {Quantum
  theory of open systems}}}\ (\bibinfo  {publisher} {Academic Press},\ \bibinfo
  {year} {1976})\BibitemShut {NoStop}%
\end{thebibliography}%


\begin{thebibliography}{99}
\bibitem[R1]{R1} Optimal transport in one-dimensional infinite chains has been
studied in \cite{MadhukarPost1977,Kumar1985,Weiss1985,plenio2008dephasing,Amir2009,Rebentrost_2009}.

\bibitem[R2]{R2}
Optimal transport in one-dimensional chains with ``exit site" has
been studied in \cite{Kaplan2017ExitSite,Kaplan2017B,CaoSilbeyWu2013}

\bibitem[R3]{R3}
The term quantum Goldilocks effect has been suggested in \cite{lloyd2011quantum}, for the idea that natural selection tends to drive quantum systems to the degree of optimal quantum coherence for transport.

\bibitem[S1]{SM1} See SM sections 1 and 2 for pedagogical presentation of the procedure for the construction of an Ohmic master equation for 2 a site system and for a chain.

\bibitem[S2]{SM2} See SM section 3 for an explicit expression for the current operator.

\bibitem[S3]{SM3} See SM section 4 for pedagogical summary regarding spreading, following \cite{Cohen1997}. 

\bibitem[S4]{SM4} See SM section 4 for technical summary of the procedure for finding the eigenvalues of a master equation with a Pauli-type dissipator. It follows \cite{EspositoGaspard2005}, but here we include additional stochastic transitions in ``parallel" to the coherent hopping, and incorporate also the bias within a first order treatment.   

\bibitem[S5]{SM5} See SM sections 5-7 for technical details regarding the procedure for finding the eigenmodes of the Ohmic master equation, including explicit expressions for the $\eta$ related terms in the Fourier representation, numerical verification for momentum thermalization, and derivation of the associated $\eta$-related correction for the diffusion coefficient.      

\bibitem[S6]{SM6} See SM section 3 for extra numerics that concerns the calculation of the current for a disordered chain, and the manifestation of the convex property. 

\bibitem[S7]{SM7} See SM section 8 for technical details regarding the derivation of the effective disorder that emerges in the reduced rate equation due to virtual coherent transitions.   

\setcounter{enumiv}{0}
\setcounter{NAT@ctr}{0}


\end{thebibliography}

\clearpage
\onecolumngrid
\pagestyle{empty}
\renewcommand\thelinenumber{}

\renewcommand{\thefigure}{S\arabic{figure}}
\setcounter{figure}{0}

\renewcommand{\theequation}{S-\arabic{equation}}
\renewcommand{\Eq}[1]{{\textcolor{blue}{Eq.}}~\!\!(\ref{#1})} 
\setcounter{equation}{0}

\Cn{{\large \bf Quantum stochastic transport along chains}} 

\Cn{{Dekel Shapira, Doron Cohen}} 

\Cn{{\large (Supplementary Material)}}


\section{Ohmic dissipator for a two site system}

Consider a two site system with Hamiltonian $\bm{H}_0$ and an Ohmic bath of temperature $T$ that induces a fluctuating force $f(t)$ of intensity $\nu$, and system-bath interaction term $-\bm{W} f(t)$. The master equation acquire a dissipation term    
\beq \label{eq:L-nu-two-sites}
\mathcal{L}^{(\text{ohmic})} \rho \ = \ 
-\dfrac{\nu}{2}\, [\bm{W}, [\bm{W}, \rho]]  
-i\dfrac{\eta}{2} [\bm{W}, \{\bm{V}, \rho\}]   
\eeq
where $\bm{V}=i[\bm{H}_0, \bm{W}]$, and $\eta=\nu/(2T)$.  
An extra noise source $\tilde{f}(t)V$ can be added in order to make the right hand side ``positive" in the Lindblad sense:
\beq 
\mathcal{L}^{(\tilde{\nu})} \rho \ = \ 
-\dfrac{\nu_{\eta}}{2}\, [\bm{V}, [\bm{V}, \rho]] 
\eeq
The ``minimal correction'' that is needed is to set $\nu_{\eta}=\nu/(4T)^2$, and then the expression can be written in the Lindblad form
\beq
\mathcal{L}^{(\text{ohmic})} \rho&  =
&\nu \left( 
\bm{F} \rho \bm{F}^{\dag} - \dfrac{1}{2} \left\{ \bm{F}^{\dag}\bm{F},\rho\right\} \right) 
- i [\bm{H}_{LS},\rho] \\
\bm{F}& =  &\bm{W} + i \frac{\eta}{2\nu} \bm{V} \\       
\bm{H}_{LS} & = & \frac{\eta}{4}[\bm{W}\bm{V}+\bm{V}\bm{W}] 
\eeq
where the Lamb-shift term $\bm{H}_{LS}$ can be absorbed into the system Hamiltonian.
For two site system with
\beq
\bm{H}_0 \ \ = \ \ -(\mathcal{E}/2) \bm{\sigma}^z - (c/2)\bm{\sigma}^x
\eeq
and coupling $\bm{W}=\bm{\sigma_x}/2$, 
one has $\bm{V} = \mathcal{E} \bm{\sigma_y}$, 
and the Lamb-shift is zero. The Lindblad generator is 
\beq
\bm{F} \ \ = \ \ 
\left(1 + \frac{\mathcal{E}}{T} \right)  \bm{\sigma}^{+} 
+ \left(1 - \frac{\mathcal{E}}{T} \right) \bm{\sigma}^{-}
\eeq
The transition rates between the sites are:  
\beq
&& w^{\pm} \ = \ \left(1 \pm \frac{\eta \mathcal{E}}{2\nu} \right)^2 \nu 
\ \approx \ (\nu  \pm  \eta\mathcal{E}) 
\\
&& \frac{w^{-}}{w^{+}} 
\ \ \approx \ \ e^{-\mathcal{E}/T} 
\eeq
A secular-like (Pauli) version of the dissipator is obtained 
by expanding $\bm{F} \rho \bm{F^{\dag}}$ 
and keeping only the Lindblad terms with $\bm{F}_{+} = \bm{\sigma}^{+}$ and $\bm{F}_{-} = \bm{\sigma}^{-}$.  
Namely,
\begin{align}
\begin{split}
\mathcal{L}^{(\text{Pauli})} \rho &= 
w^{+} \left(  \bm{\sigma}^{+}\rho \bm{\sigma}^{-} -\frac{1}{2}\{ \bm{\sigma}^{-}\bm{\sigma}^{+}, \rho \}\right)
+ w^{-} \left( \bm{\sigma}^{-}\rho \bm{\sigma}^{+} -\frac{1}{2}\{ \bm{\sigma}^{+}\bm{\sigma}^{-}, \rho \}\right)
-\dfrac{\gamma}{4}\, [\bm{\sigma}^z, [\bm{\sigma}^z, \rho]]  
\end{split}
\end{align}
where the last term represents excess noise due to noisy detuning (see below).
The mixed terms that have been omitted affect only the decoherence of the off-diagonal terms, 
and not the rate of transitions between sites. In the Bloch-vector representation 
the precessing component of the ``spin" decays only in the $y$ direction in the Ohmic version, 
and isotropically in the Pauli version.  In some sense the dissipation in the Pauli version 
assumes two independent baths at each bond. 
If we assume that the detuning $\mathcal{E}$ is fluctuating with intensity $\nu_{\gamma} = (\gamma/2)$, so that 
$(\mathcal{E}/2) \rightarrow (\mathcal{E}/2) + f(t)$,
then an additional $\mathcal{L}$ term appears, that has the form of \Eq{eq:L-nu-two-sites} 
with the substitution $\bm{W}:=\bm{\sigma}^z$, and $\bm{V}:=-c\bm{\sigma}^y$.

\newpage
\section{Ohmic dissipator for a chain}

The Hamiltonian of the chain is
\beq
\bm{H}^{(c)} \ = \ U(\bm{x}) - \frac{c}{2}(\bm{D}+\bm{D}^{\dag}) \ =\ -\mathcal{E} \bm{{x}} - c\cos(\bm{p}) 
\eeq
where $\bm{D}=\sum_x \bm{D}_x $ is the displacement operator, and $\bm{D}_x = \kb{x+1}{x}$. 
In general the field ${\mathcal{E}_x = -\left( U(x{+}1)-U(x) \right)}$, 
as well as the hopping frequencies ($c$), and temperatures might be non-uniform.  
The interaction with a bath-source that induces non-coherent transitions at a given bond is obtained by the replacement ${(c/2)\mapsto (c/2)+ f(t)}$.
The baths of different bonds are uncorrelated.  
Accordingly the dissipation term in the Master equation takes the form 
\beq \label{eq:dro-dt-OME}
\mathcal{L}^{(\text{ohmic})} \rho = -\sum_{x} \left(
\dfrac{\nu}{2} [\bm{W}_x, [\bm{W}_x, \rho]] 
+ \dfrac{\eta}{2}\, i [\bm{W}_x, \{\bm{V}_x, \rho\}]  
\right) 
\eeq
where the coupling to the baths is via the operators 
\begin{align}
 \label{eq:ome-terms}
\bm{W}_x &= \left(\bm{D}_x + \bm{D}_x^{\dag} \right) \\
\bm{V}_x &= i[\bm{H}^{(c)}, \bm{W}_x] =
i \mathcal{E}_x \left(\bm{D}_x^{\dag} - \bm{D}_x\right) - 
i \dfrac{c}{2} \left[   \left( \bm{D}_{x+1} \bm{D}_x - \bm{D}_x \bm{D}_{x-1} \right)  - h.c \right] 
\end{align}
And the Lindblad correction term:
\beq
\mathcal{L}^{(\tilde{\nu})} \rho \ = \ - \dfrac{\nu_{\eta}}{2}  \sum_{x}  [\bm{V}_x, [\bm{V}_x, \rho]]
\eeq
with intensity $\nu_{\eta} = \nu/(4 T)^2$.
Such term has negligible effect in the high temperature regime (${\eta < 1}$). 
Optionally we can add terms that reflect fluctuations of the field. 
At a given bond it is obtained by the replacement ${ U(\bm{x}) \mapsto U(\bm{x}) + \tilde{f}(t)}$, 
where $\tilde{f}(t)$ represents fluctuations of intensity $\gamma$. 
The implied coupling operators are 
\beq 
\bm{W}^{(S)}_x &=& \bm{Q}_x \\
\bm{V}^{(S)}_x &=& i[\bm{H}^{(c)}, \bm{W}^{(D)}_x] \ = \ i (c/2) \left[ \bm{D}_{x-1}^{\dag} - \bm{D}_{x-1} - \left( \bm{D}_x^\dag - \bm{D}_{x} \right)\right]
\eeq

\section{Expression for the current}
\label{sec:current}

For generality of the treatment we allow the temperature to be bond dependent, 
then $\eta \rightarrow \eta_x$ so that the
Lindblad generators are $\bm{F}_x =  \bm{W}_x + i (\eta_x/2\nu) \bm{V}_x$, and 
\beq
&& \bm{H} = \bm{H}^{(c)} + \sum_{x}\dfrac{\eta_x}{4} \{ \bm{W}_x, \bm{V}_x \}
\eeq        
The time dependence of an expectation value is given by the adjoint equation: 
\beq \label{eq:Q-dot}
\begin{split}
\frac{d}{dt} \braket{\bm{Q}} 
\ = \ \trc\left[ \bm{Q} \frac{d}{dt}\rho \right]
\ = \  \trc\left[ \bm{Q} \mathcal{L} \rho \right]
\ = \  \trc\left[ (\mathcal{L}^{\dag}  \bm{Q})  \rho \right]
\ = \  \braket{\mathcal{L}^{\dag} \bm{Q}} 
\end{split}
\eeq
where
\beq
\mathcal{L}^{\bdag} \bm{Q} \ = \
i[\bm{H}, \bm{Q}] +
\nu \sum_x \left( \bm{F_x}^{\dag} {\bm{Q}} \bm{F_x} -  \dfrac{1}{2} \{  \bm{F_x}^{\dag} \bm{F_x}, {\bm{Q}} \}  \right) 
\eeq
Partitioning the system at the $n$-th bond,  
the current flowing from left to right is defined by $I = \dot{\avg{\bm{Q}}}$, with 
\beq
\bm{Q} \ \ = \ \ \sum_{x>n} \kb{x}{x}
\eeq
We note that although the original Hamiltonian allows only near-neighbor hopping, 
the master equation allows also ``double hopping'' due to the $\bm{V}$ terms.
Accordingly the expression for the current operator has several non-trivial terms.
Applying \Eq{eq:Q-dot} the current is:
\begin{align} \label{eq:current-across-interface}
I & \ = \ \vec{I} - \lvec{I} - c\,\im[\rho_n{(1)}] + I_{\eta^2}^{(0)} + I_{\eta^2}^{(1)}  \\
\vec{I} & \ = \ w^+_n p_n  \label{eq:current-interface}  -\dfrac{c \eta_n }{2} \re[\rho_{n-1}{(1)}]  \\
\lvec{I} & \ = \ w^{-}_n p_{n+1} -\dfrac{c \eta_n }{2} \re[\rho_{n+1}{(1)}]  \label{eq:I-right} \\
I_{\eta^2}^{(0)} & \ = \  \dfrac{\mathcal{E}^{2} \eta_n^2}{4 \nu} [p_n - p_{n+1}] +\sum_{i=0,1} \dfrac{c^{2}}{16 \nu}
\left( \eta_{n-i}^2 + \eta_{n+1-i}^2 \right) \left( p_{n-i} - p_{n+2-i} \right) \\
I_{\eta^2}^{(1)} & \ = \ 
- \dfrac{c \mathcal{E} }{8 \nu} \re \left[2 \eta_n^2 \left(\rho_{n -1}{(1)} + \rho_{n+1}{(1)}  \right) 
- \left( \eta_{n-1}^2 + \eta_{n+1}^2 \right) \rho_n{(1)}\right]
\end{align}
where the extra $I_{\eta^2}$ terms are of order $\eta^2$,
and are negligible for the NESS current.
If the field $\mathcal{E}$ is non-uniform, then one needs
to make the replacement $\eta_n \mathcal{E} \rightarrow \eta_n \mathcal{E}_n$.
Disregarding $I_{\eta^2}$ the distinct elements of the current are 
coherent hopping, stochastic hopping and stochastic-assisted coherent hopping.
These are pictured in \Fig{fig:pn-transitions}.

\sect{Current in disordered system}
In \Fig{fig:I-vs-corr} we display results for the NESS current, 
calculated for a disordered sample. If the spatial correlation scale 
of the disorder is large, the ring can be regarded as composed   
of several segments connected in series. Then the analytical 
estimate for the current would be  
\beq \label{eq:I-estimate-disorder}
I \ \ = \ \ \left[ \sum_x \frac{1}{v(\mathcal{E}_x)} \right]^{-1}
\eeq
which reduce to $I=(1/L)v(\mathcal{E})$ for a uniform field. 
The function $v(\mathcal{E})$ is provided by \Eq{eq:I}, 
and the above analytical estimate implies what we call {\em convex property}. 
Using this formula we can explain why disorder 
can lead to increase of the current as in \Fig{f1}.  
The accuracy of this formula, that assumes a large spatial correlation scale, 
is tested against the correlation scale in \Fig{fig:I-vs-corr}.

\begin{figure}[h]
\centering
\includegraphics[width=5cm]{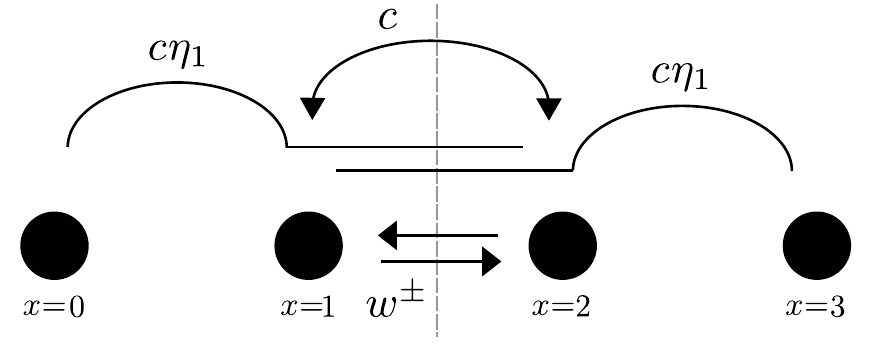}  
\caption{
Diagrammatic representation of the terms that contributes to the total current that flows via a section that is indicated by a vertical dashed line, here via the bond that connects sites $x{=}1$ and $x{=}2$.
Straight lines denotes the role played by the stochastic transitions, 
while semi-circle segments are related to coherent hopping. The latter are of the form $c \eta_x \rho_x{(1)}$.
\label{fig:pn-transitions} }
\end{figure}

\begin{figure}[h]
\centering
\includegraphics[width=7cm]{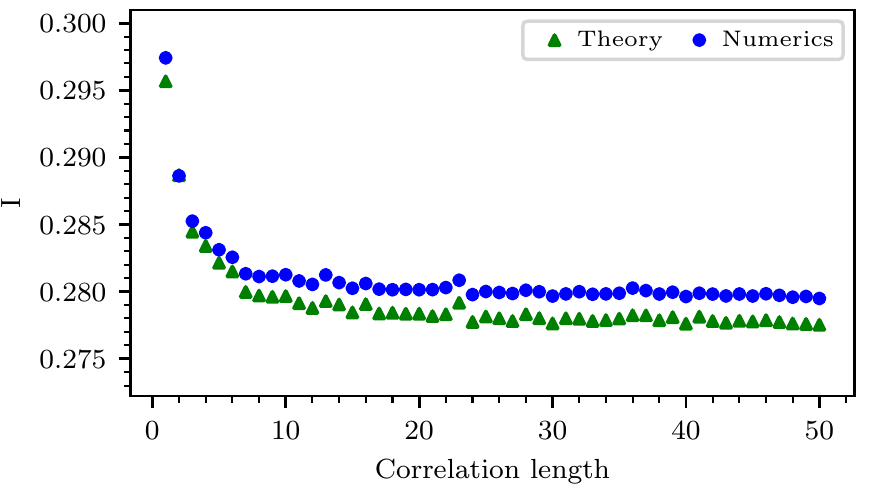}~
\includegraphics[width=7cm]{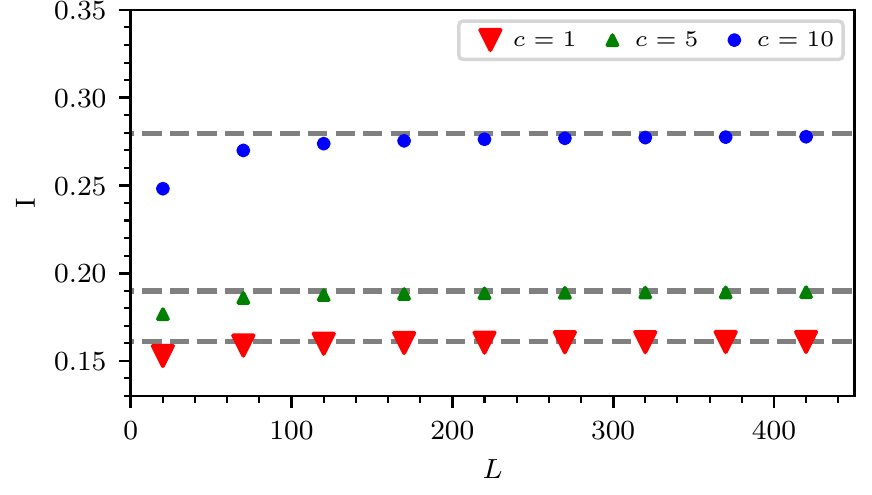}
\caption{\label{fig:I-vs-corr}
The NESS current as a function of the correlation length of the disorder. 
The parameters are $\mathcal{E}{=}8$, $\sigma_{\mathcal{E}}{=}6$, $\eta{=}0.01$ and $c{=}10$.
The correlated disorder is obtained by a convolution of the uncorrelated 
disorder (correlation length equals unity) with a box-shaped kernel. 
For each sample the numerical result for an $L{=}500$ ring is compared 
with the theoretical estimate \Eq{eq:I-estimate-disorder}.
The two are expected to coincide for large correlation length.  
The small difference that remains is a finite size effect
due to our treatment of the boundary conditions (see main text).
On the right panel we demonstrate the $L$-dependence 
of this difference for a clean system ($\mathcal{E}$ and $\eta$ are the same). 
The gray dashed lines are the theoretical values provided by \Eq{eq:I}.
}
\end{figure}

\newpage
\section{Spreading}

Without coherent hopping ($c{=}0$) the Ohmic/Pauli dynamics of the on-site probabilities $p_x$ 
decouple from the decay of the off-diagonal terms.  
We get two distinct sets of modes: the stochastic-like relaxation modes and the off-diagonal decoherence modes.
In the Pauli approximation the latter decay with the {\em same} rate ${\gamma_0=\gamma+w^{+}+w^{-}}$. 
The stochastic transitions affect only the stochastic-like relaxation modes.
Starting with a wavepacket of variance $\text{Var}(R)$ =  $\sigma_0^2$, and momentum centered around $k_0$,
we get in the Wigner representation  
\beq
\rho_{w}(R,P) \ \ = \ \ e^{-\gamma_0 t} \, \left[ G^c(R,P)-G^0(R,P) \right] \ + \ G^t(R,P)  
\eeq
where 
\beq
G^c(R,P) &=& \dfrac{2}{L}\exp \left( -\dfrac{1}{2}\dfrac{R^2}{\sigma_0^2} -(P-k_0(t))^2 \sigma^2 \right), 
\ \ \ \ \ \ \ k_0(t)=k_0+\mathcal{E}t 
\\
G^t(R,P) &=& \dfrac{1}{\sqrt{2 \pi}} \dfrac{1}{\sqrt{\sigma_0^2 + 2 Dt}} \exp \left( -\dfrac{1}{2}\dfrac{(R - vt)^2}{\sigma_0^2 + 2Dt} \right)
\eeq
with drift velocity $v=(w^{+} - w^{-})$ and diffusion coefficient $D=(w^{+} + w^{-})/2$. 

Let us add coherent hopping in a very naive way: we use the Pauli dissipator, 
and merely set ${c\ne 0}$ in the Hamiltonian. It is a naive procedure because the use 
of the Pauli dissipator cannot be justified anymore (we have to use the Ohmic dissipator).    
With this simplification the explicit form of the Lindblad operator of the $q$ block is:
\begin{align}\label{eq:sm-secular}
\mathcal{L}^{(q)} \ \ = \ \ - \gamma_0   
\ +  \gamma_q  \kb{0}{0} 
\ -i \mathcal{E} \sum_r \ket{r}r\bra{r}
\ - c \sin(q/2) \left[e^{iq/2}\mathcal{D}_{\perp}-e^{-iq/2}\mathcal{D}_{\perp}^{\dag}\right]
\end{align}
with $\gamma_q =  \gamma + w^{+} e^{-iq} + w^{-} e^{iq}$. 
This can be regarded as simplified version of \Eq{eq:sm-L-ohmic-1} below. 
The diagonalization of $\mathcal{L}$ is straightforward for zero bias. 
The phases $\exp(\pm iq/2)$ can be gauged to some distant $r$, 
and $\mathcal{L}$ becomes like the Hamiltonian of a tight-binding model 
with a barrier at the origin. The lowest eigenmodes are decaying exponents 
$\psi(r) \sim \exp(-\alpha|r|)$, with $\re(\alpha)>0$.
These modes correspond to the stochastic-like relaxation modes. 
Matching the boundary conditions at $r=0$, one finds the eigenvalues
\beq
\lambda_{q,0} \ \ = \ \ \gamma_0 - \sqrt{\gamma_q^2 - 4 c^2 \sin^2{(q/2)}} 
\ \ \equiv \ \ i v q + D q^2 + O(q^3) 
\eeq
From which expressions for $v$ and $D$ can be derived.
The result for $D$ is similar (but not identical) to the correct 
result in the main text. Namely, up to a prefactor it reproduces 
the $\mathcal{E}{=}0$ Drude term.

\section{Bloch representation of the Ohmic master equation}
\label{sec:OME-chain}

For a clean system, and neglecting the $\eta^2$ contribution,
the generator of the master equation is written as a sum of several terms.
Here we shall provide explicit expressions of the $q$ block 
of the super-matrix in the Bloch representation:  
\beq \label{eq:sm-L-ohmic-1}
\mathcal{L}^{(q)} \ = \ 
c\mathcal{L}^{(c)}
+\mathcal{E} \mathcal{L}^{(\mathcal{E})}
+\nu \mathcal{L}^{(\nu)}
+\eta c \mathcal{L}^{(\tilde{c})}
+\eta \mathcal{E} \mathcal{L}^{(\tilde{\mathcal{E}})}
+\nu_{\eta} \mathcal{L}^{(\tilde{\nu})} 
\eeq
We define operators 
\beq
R \ &=& \ \sum_r \ket{r}r\bra{r} \\  
\mathcal{D}_{\perp} &=& \sum_r \ket{r+1}\bra{r}
\eeq
After gauge transformation ${\ket{r} \rightarrow e^{-iq r/2}\ket{r}}$ we obtain 
\beq  \label{eq:sm-L-ohmic-2}
\mathcal{L}^{(c)} &=& \sin(q/2) [\mathcal{D}_{\perp}^{\dag} -\mathcal{D}_{\perp}] \\ 
\mathcal{L}^{(\mathcal{E})} &=& -iR  \\
\mathcal{L}^{(\nu)} &=& -2 + 2\cos(q) \kb{0}{0} + \left(\kb{1}{{-1}} + \kb{{-1}}{1}  \right) \\
\mathcal{L}^{(\tilde{c})} &=&  \frac{1}{2}\cos{(q/2)} [\mathcal{D}_{\perp}+\mathcal{D}_{\perp}^{\dag}]  
+ \dfrac{1}{2} \cos(3q/2) \left[  \kb{\pm 1}{0} - \kb{0}{\pm 1} \right]
+ \dfrac{1}{2}\cos(q/2) \left[ \kb{{\mp 2}}{\pm 1} - \kb{\pm 1}{{\mp 2}} \right] \\
\mathcal{L}^{(\tilde{\mathcal{E}})} &=& -2i \sin{(q)} \kb{0}{0}
\eeq
Note that this expression is not $2 \pi$ periodic,
since we ignore the accumulated phase which arise in the gauge procedure.
The gauge in the above procedure is equivalent to redefinition of the $r$ coordinate 
such that $x$ and $r$ become orthogonal (skewing the $r$ axis in \Fig{fig:schematic-transitions-all} by 45 degrees).

\newpage
\section{Eigenmodes of the Ohmic master equation}

\sect{Infinite temperature eigen-modes}
For infinite temperature ($\eta=0$), the eigenvalues
of the $q=0$ block are:
\beq 
\lambda_{q{=}0,0}  &=& 0 \ \text{(NESS)} \\
\lambda_{q{=}0,\pm}  &=&  2\nu \pm \sqrt{\nu^2 - \mathcal{E}^2} \\
\lambda_{q{=}0,s}  &=& 2\nu + i\mathcal{E} s, \ \ (s = \pm 2, \pm 3,...) 
\eeq
Considering the $q$ dependence of the eigenvalues we get several bands. Our interest below is in the lowest band ($\lambda_{q,s=0}$),
which determines the long time spreading.     
For this calculation one needs the eigen-modes corresponding to the above eigenvalues. These are given by:
\beq
&& \ket{\lambda_{q{=}0,s}} = \ket{r=s}, \ \ (s =0, \pm 2, \pm 3,...)  \\
&& \ket{\lambda_{q{=}0,\pm}} \equiv \ket{\pm} \ = \ \alpha_{\pm} \ket{1} + \ket{-1} \ \ \  \text{(unnormalized)} \\
&&  \alpha_{\pm} = - i\left(\dfrac{\mathcal{E}}{\nu}\right) \mp \sqrt{1 - \left(\dfrac{\mathcal{E}}{\nu}\right)^2}  \label{eq:plus-minus-basis}
\eeq

\sect{NESS at finite temperature}
We can find the NESS, which is the zero mode $\ket{\lambda_{0,0}=0}$, 
and calculate from it both the momentum distribution and the current.

Setting ${q=0}$, and considering linear order in $\eta$,
the NESS is obtained by first order perturbation for the $\ket{\lambda_{q=0,0}} = \ket{r=0}$ state. 
See \Fig{fig:ohmic-q-r-lattice}. Putting $V \equiv \eta c \mathcal{L}^{(\bar{c})}$ as the perturbation,
one get:
\beq
&& \ket{\text{NESS}} \ \ = \ \ 
\ket{0} + \dfrac{\BraKet{\tilde{+}}{V}{0} }{\lambda_+} \ket{+} 
+ \dfrac{\BraKet{\tilde{-}}{V}{0}}{\lambda_-}\ket{-} \ \ = \ \  
\ket{0} + \alpha_0 \ket{1} + \alpha_0^{*} \ket{-1}  \\
&& \alpha_0 = \dfrac{3\nu - i\mathcal{E}}{3 \nu^2 + \mathcal{E}^2} \, \eta c 
\eeq
where the left eigenvectors are given by:
\begin{align}
\bra{\tilde{\pm}} \ \ = \ \  
\left( \dfrac{\alpha_{\pm}}{\alpha_{\mp}} - 1 \right)^{-1}
\left[ \bra{1} \dfrac{1}{\alpha_{\mp}} - \bra{-1}  \right],  
\end{align}
Reverting back from the Bloch basis of $\rho(r;q)$ to the position basis, 
namely $\ket{r;q} := {L^{-\frac{1}{2}}}\sum_{x} \kb{x}{x\!+\!r} e^{i q x}$,
the normalized steady state matrix $\rho$ is:
\beq \label{eq:rho-steady-state}
\rho^{(\text{NESS})} = \dfrac{1}{L}\left( \id +   \sum_{x} \alpha_0 \kb{x}{x{+1}}  + h.c.\right) 
\ = \ \dfrac{1}{L}\left( \id + \alpha_0 e^{{+}ip} + \alpha_0^{*} e^{-ip}\right)
\eeq

\sect{The momentum distribution}
Using \Eq{eq:rho-steady-state} we obtain the steady state momentum distribution:
\beq \label{eq:sm-pk}
p(k) = \rho_{kk} = \dfrac{1}{L} \left( 1 + 2 \re(\alpha_0 e^{+ik}) \right) =
\dfrac{1}{L} + \dfrac{1}{L} \dfrac{2 \eta c}{3 \nu^2 + \mathcal{E}^2} 
\left( 3 \nu \cos{(k)} +  \mathcal{E} \sin{(k)}\right) 
\eeq
For $\mathcal{E}=0$, the momentum distribution is canonical, see \Fig{fig:rho-p-canonical}.
The above result is indeed consistent with the canonical distribution to linear order in $\beta=1/T$. 
The drift velocity can be deduced by calculating the NESS current using \Eq{eq:current-across-interface}. 
The current over the bond~$n$, to first order in $\eta$ is:
\beq
I_n \ \ = \ \ \frac{1}{L} \left(  (w^+_n - w^-_n) - c\, \im(\alpha_0)  \right) 
\ \ = \ \ \frac{1}{L}  \left[ 1 + \frac{c^2 }{6\nu^2 + 2\mathcal{E}^2} \right] 2\eta \mathcal{E}
\eeq
We note that although the expression for the current \Eq{eq:current-across-interface} is complicated, the final NESS current
is composed of the usual stochastic-current, and the usual coherent-current.

\newpage
\section{Diffusion at finite temperature}
 
Here we provide the calculation of $D$ to second order in $\eta$. 
We have to expand $\lambda_{q,0}$ to second order in $q$.
Inspecting the gauged Lindblad operator in \Eq{eq:sm-L-ohmic-1}, 
one observes (see diagram of \Fig{fig:ohmic-q-r-lattice}) 
that up to order $q^2$ and $\eta^2$ it is enough to diagonalize the five sites $|r| \le 2$, 
keeping the $q^2$ and the $\eta^2$ corrections. 
This can be done using perturbation theory, 
or optionally using \textit{Mathematica} for a direct diagonlization 
and then expand the result in powers of $q$ and $\eta$. 
Either way one get:
\beq
\lambda_{q,0} &=& iv q + D q^2
\\ \label{eq:drift-ohmic} 
v &=&  \left[ 1 + \frac{c^2 }{6\nu^2 + 2\mathcal{E}^2} \right] 2\eta \mathcal{E} 
\\  \label{eDeta}
D &=&  \left[ 1 + \frac{c^2 }{6\nu^2 + 2\mathcal{E}^2} \right] \nu
- \left[ 
\dfrac{ \left(9 \nu ^2+11 \mathcal{E} ^2\right)}{(\mathcal{E}^2 + 3 \nu^2)^2}
+ \dfrac{\left(15 \mathcal{E}^2+13 \nu^2\right) (c\mathcal{E})^2}{4 (\mathcal{E}^2 + \nu^2) (\mathcal{E}^2 + 3 \nu^2)^3}
\right] (\eta c)^2 \nu  
\eeq
Setting $\mathcal{E}=0$ in the expression for $D$, we find that the $\eta^2$ correction 
in \Eq{eDeta} can be absorbed into the first term via 
the replacement ${c^2 \mapsto [1-6 \eta^2]c^2 }$.  
Note that this correction is based on the Ohmic dissipator 
without the additional Lindblad term that is added for the purpose of positivity.

\begin{figure}[h]
\centering
\includegraphics[width=7cm]{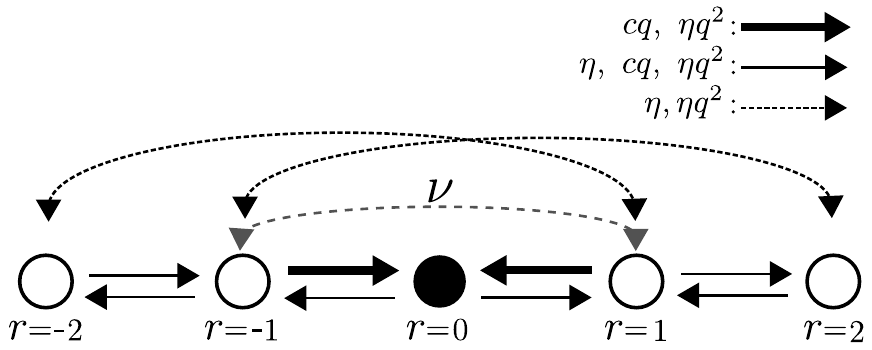}
\caption{\label{fig:ohmic-q-r-lattice}
Diagrammatic representation of the couplings in the reduced tight binding model (in~$r$), 
that is used in order to determine the eigenvalues $\lambda_{q,s}$ for a given Bloch momentum~$q$. 
Different orders of $q$ and $\eta$ are indicated by the different arrows.
The formation of the $\ket{\lambda_{q{=}0,\pm}}$ eigenmodes is due to the dashed $\nu$ coupling.
Up to order $q^2$ and $\eta$ it is enough to consider second order perturbation theory involving ${r=-1,0,1}$. 
For $\eta^2$ corrections one needs to include also ${r=\pm 2}$. }
\end{figure}

\begin{figure}[h]
\centering
\includegraphics[width=7cm]{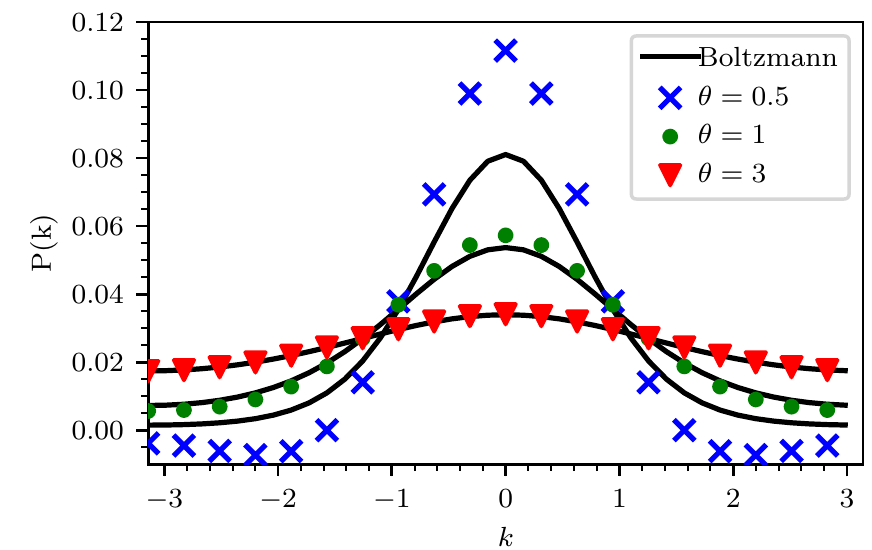}
\includegraphics[width=7cm]{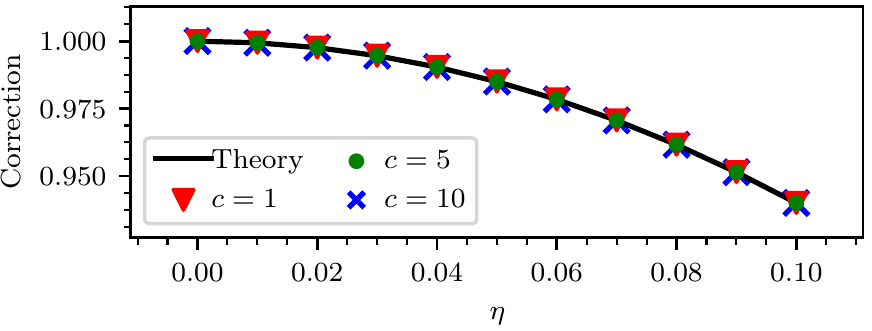}
\caption{\label{fig:rho-p-canonical}
The steady state and the diffusion coefficient at zero bias.  
The parameters are ${\nu=1}$ and ${\eta=0.01}$.
\textbf{(a)} Numerically determined momentum distribution (without the Lindblad correction)
compared with the Boltzmann distribution ${p(k) \propto \exp \left[ \theta^{-1} \cos{(k)} \right]}$.
The distribution is plotted for a few values of $\theta$.
For small $\theta$ the distribution is no longer Boltzmann-like and its 
tails become negative. The Ohmic master equation is no longer valid in this regime.
The calculation is for $L=40$, and for clarity only 20 data points are presented.
\textbf{(b)} Numerical verification of the $\eta^2$ correction to $D$.
The analytically predicted correction factor $[1 - 6 \eta^2]$ is plotted as a solid line.
The numerical data points are $[6 \nu/c^2](D-\nu)$.
}
\end{figure}

\newpage
\section{Effective disorder}

For $c{=}0$ the off-diagonal terms of $\rho$ decouple from the diagonal,
and the relaxation spectrum is the same as that of the stochastic model.
For small $c$, we keep only the three-central diagonals ($r \le 1$), 
thus ignoring the couplings to higher bands.
We also ignore transitions of the order $c \eta$. 
The  ${r=\pm 1}$ space can be eliminated, in the price of
getting a $\lambda$-dependent transition rate matrix 
$H_{\text{eff}}(\lambda) = H_0 + W'G(\lambda)W$,
where $H_0$ includes the transitions along $r{=}0$, 
and $G(\lambda)$ is a resolvent operator that describes the dynamics within the excluded diagonal, 
while the $W$-s include the couplings between the $r{=}0$ elements and the excluded $r=\pm 1$ elements.  
See diagram of the couplings in \Fig{fig:3-channel}. 
The effective matrix $H_{\text{eff}}(\lambda)$ is non-hermitian 
due to the asymmetry of the transitions.
It is a probability conserving tight-binding operator,  
but with rates that can be negative. 
For the forward and backward hopping rates in the $n$-th bond 
we get ${w^{\pm}_n  = \nu + \nu_n \pm \eta \mathcal{E}_n}$, where 
\beq
\nu_n \ \ = \ \
\left( \dfrac{c}{2} \right)^2
\left( G_{11} + G_{22} - G_{12} - G_{21}\right) 
\ \ = \ \ \dfrac{c^{2}}{2} \dfrac{\nu - \lambda}{(2 \nu - \lambda)^2 + \mathcal{E}^2_n - \nu^2}
\eeq
where $G = -(\lambda + L_n)^{-1}$ is a $2 \times 2$ matrix 
which is defined in terms of $L_n = -2 \nu - i \mathcal{E}_n \bm{\sigma}_z + \nu \bm{\sigma}_x$ 
within the subspace that is spanned by the super-vectors $\kb{n}{n+1}$ and $\kb{n+1}{n}$. 
The eigenvectors of this matrix are the $\ket{\pm}$ of \Eq{eq:plus-minus-basis} 
with $\mathcal{E} \rightarrow \mathcal{E}_n $.



In order to estimate the effective disorder, we proceed as outlined in the main text. 
For high-temperatures one obtains from \Eq{eq:eff} approximations for the $w_n$ 
and for the stochastic field, namely, ${w_n \approx (\nu + \nu_n)}$, 
and ${\tilde{\mathcal{E}}_n \approx \eta \mathcal{E} (\nu + \nu_n)^{-1}}$.
We define an associated hermitian matrix $\tilde{H}$, 
that has the same matrix elements as $H_{\text{eff}}(\lambda)$, 
but with $\tilde{\mathcal{E}}_n=0$ in the off diagonal elements. 
The eigenvalues of $\tilde{H}$ are real, with some inverse localization length $\kappa(\lambda)$.
Ignoring the diagonal disorder that arises due to non-uniform field $\tilde{\mathcal{E}}_n$,
the localization length of eigenvalues near $\lambda$ are roughly given by 
[\href{http://dx.doi.org/%2010.1103/PhysRevE.93.062138}{Weinberg, de Leeuw, Kottos, Cohen, Phys. Rev. E 93, 062138 (2016)}]:
\begin{align}
\kappa(\lambda) \ \ \approx \ \  
\dfrac{1}{4} \left( \dfrac{\sigma_{\perp}}{\nu}  \right)^2 \dfrac{\lambda}{\nu}
\end{align}
with $\sigma^2_{\perp} = \var(w_n)$. 
This, as explained in the main text, determines whether 
the eigenvalues of $H_{\text{eff}}(\lambda)$ will turn complex.
We focus on representative region around  $\lambda=2\nu$ in the center of the spectrum.  
Around this point, for small  disorder, one obtains:
\begin{align}
&w_n \ \approx \ \nu \dfrac{c^{2}}{2 (\nu^2 - \mathcal{E}^2)} \left( 1 + B \delta_n + C \delta_n^2 \right)\\
&B  = \frac{2 \mathcal{E}}{(\nu^2 - \mathcal{E}^2)},  
\ \ \ \ \ \ \ \ \ \ 
C = \dfrac{\nu^{2} + 3 \mathcal{E}^2}{(\nu^2 - \mathcal{E}^2)^2}
\end{align}
with $\delta_n \equiv \mathcal{E}_n - \mathcal{E}$ 
that are randomly distributed within $[-\sigma_{\mathcal{E}},\sigma_{\mathcal{E}}]$. 
Consequently we get the estimate
\begin{align} \nonumber
&\sigma^2_{\perp} \ = \ \var(w_n) \ \approx \  
\left( \dfrac{c^{2} \nu}{2 (\nu^2 - \mathcal{E}^2)} \right)^2
\left(B^2\, \var(\delta) + C^2\, \var(\delta^2)  \right) 
=
\left(\dfrac{c^{2}\nu}{2 (\nu^2 - \mathcal{E}^2)} \right)^2
\left(B^2 (\sigma^2/3) + C^2 (4 \sigma^4/45)  \right) 
\end{align}

\begin{figure}[h]
\centering
\includegraphics[height=35mm]{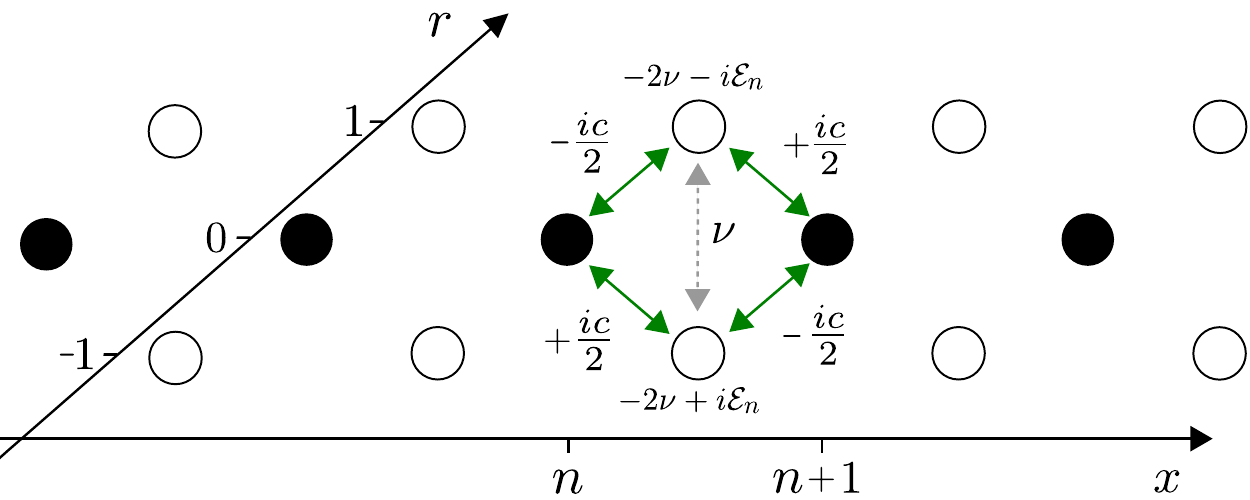}
\caption{\label{fig:3-channel} 
Diagrammatic representation of the couplings in the 3 band approximation. 
Along the main diagonal (filled circles) we have asymmetric stochastic transitions (not indicated).
Those are coupled to the coherences (empty circles) due to the 
$c$-related terms that are packed into $W$ and $W'$ matrices.
The non-Pauli $\nu$~coupling and the on-site ``energies'' at the $|r|=1$ sites 
constitute the $L_n$ operator, which determines the resolvent $G(\lambda)$. 
}
\end{figure}

\clearpage
\end{document}